\documentstyle[epsf,amssymb]{article}

\begin{document}
\begin{center}

{\Large {\bf Quantum critical phenomena of unconventional superconductors: 
U(1) gauge model of link Cooper pair }}\\
\vspace{1cm}
{\Large Kenji Sawamura, Ikuo Ichinose, and Yuki Moribe} \\
\vspace{1cm}
{Department of Applied Physics,
Nagoya Institute of Technology, Nagoya, 466-8555 Japan 
} 

{\bf abstract}   
\end{center}
In this paper we shall study quantum critical behavior of lattice 
model of unconventional superconductors (SC) that was proposed 
in the previous papers.
In this model, the Cooper-pair (CP) field is defined on lattice links in order
to describe $d$-wave SC.
The CP field can be regarded as a U(1) lattice gauge field,
and the SC phase transition takes place as 
a result of the phase coherence of the CP field.
Effects of the long-range Coulomb interactions between the CP's 
and fluctuations of the electromagnetic field are taken into account.
We investigate the phase structure of the model and the critical behavior
by means of the Monte Carlo simulations.
We find that the parameter, which controls the fluxes (vortices) of the
CP, strongly influences the phase structure.
In three-dimensional case, the model has rich phase structure.
In particular there is a ``monopole proliferation" phase transition besides 
the SC phase transition.
Depending on the parameters, this transition exists within the SC phase
or takes place simultaneously with the SC transition.
This new type of transition is relevant for unconventional SC's with strong
spatial three-dimensionality and to be observed by experiments.

\newpage

\section{Introduction}
Some of the experiments on the high-$T_c$ materials indicate that the
superconducting (SC) phase transition at vanishing temperature ($T$)
is a {\em quantum} phase transition\cite{exp} and it belongs to 
the universality 
class of the four-dimensional (4D) XY model\cite{XY}.
On the other hand for constructing a Ginzburg-Landau (GL) theory of the 
$d$-wave
SC phase transition, the Cooper-pair (CP) field must be put on lattice 
links instead of
site because its condensation has different signature depending on the 
direction of the CP (see later discussion).
It is also suggested that the SC transition at $T=0$ is a 
{\em phase decoherent} 
phase transition and the SC transition of the high-$T_c$ cuprates
is more like the Bose-Einstein condensation than
the BCS transition for the coherent length of the high-$T_c$ SC's
is very short compared to the ordinary metallic SC's.

In the previous papers\cite{UV1,UV2,UV3} from the above point of view, we have 
introduced a CP field $V_{x,i}$
\begin{equation}
V_{x,i} \sim \langle \psi_{x,\uparrow}\psi_{x+i,\downarrow}-
             \psi_{x,\downarrow}\psi_{x+i,\uparrow}\rangle
\label{Vxi}
\end{equation}
where $\psi_{x, \sigma}(\sigma=\uparrow, \downarrow)$ is the electron
at site $x$ with spin $\sigma$ and $i=1,\cdots, d$ is the spatial
direction index (it also denotes the unite vector).
As we study a quantum phase transition at $T=0$, the GL theory is defined
on the $d+1$ ($d$-spatial and $1$-time) dimensional hypercubic lattice.
The CP field $V_{x,i}$ has electric charges at $x$ and $x+i$ and it also
interacts with the transverse electromagnetic (EM) vector potential $A_{x,i}$.

For the ordinary $s$-wave SC, the GL theory defined on lattice plays an
important role for study on the critical phenomena.
There the Cooper pair 
$\phi_x\sim \langle \psi_{x\uparrow}\psi_{x\downarrow}-
             \psi_{x\downarrow}\psi_{x\uparrow}\rangle$ 
is put on lattice site, and its condensation induces the SC phase.
This GL theory is called gauge-Higgs model in the elementary particle physics
and plays an important role in the unified theories.
Recently the gauge-Higgs models have renewed interests and have been 
studied intensively.
In particular, dynamics of multi-flavor Higgs models and their phase structure,
order of Higgs phase transitions, effect of the Berry's phase term, etc.
are important topics.
(See Refs.\cite{Higgs1,Higgs2}.)
These studies are related to multi-gap SC's, deconfined critical points in
strongly-correlated electron systems, superfluid and SC in liquid hydrogen
at low-$T$ and in high pressure, etc.
We expect that the gauge-Higgs model with the link Higgs field $V_{x,i}$
also plays an important role in various fields of physics, and
to clarify its dynamics gives useful insight to critical phenomena.
In the present paper, we focus on its application to the unconventional (UC)
SC.

\begin{figure}[htbp]
\begin{center}
\epsfxsize=8cm
\epsffile{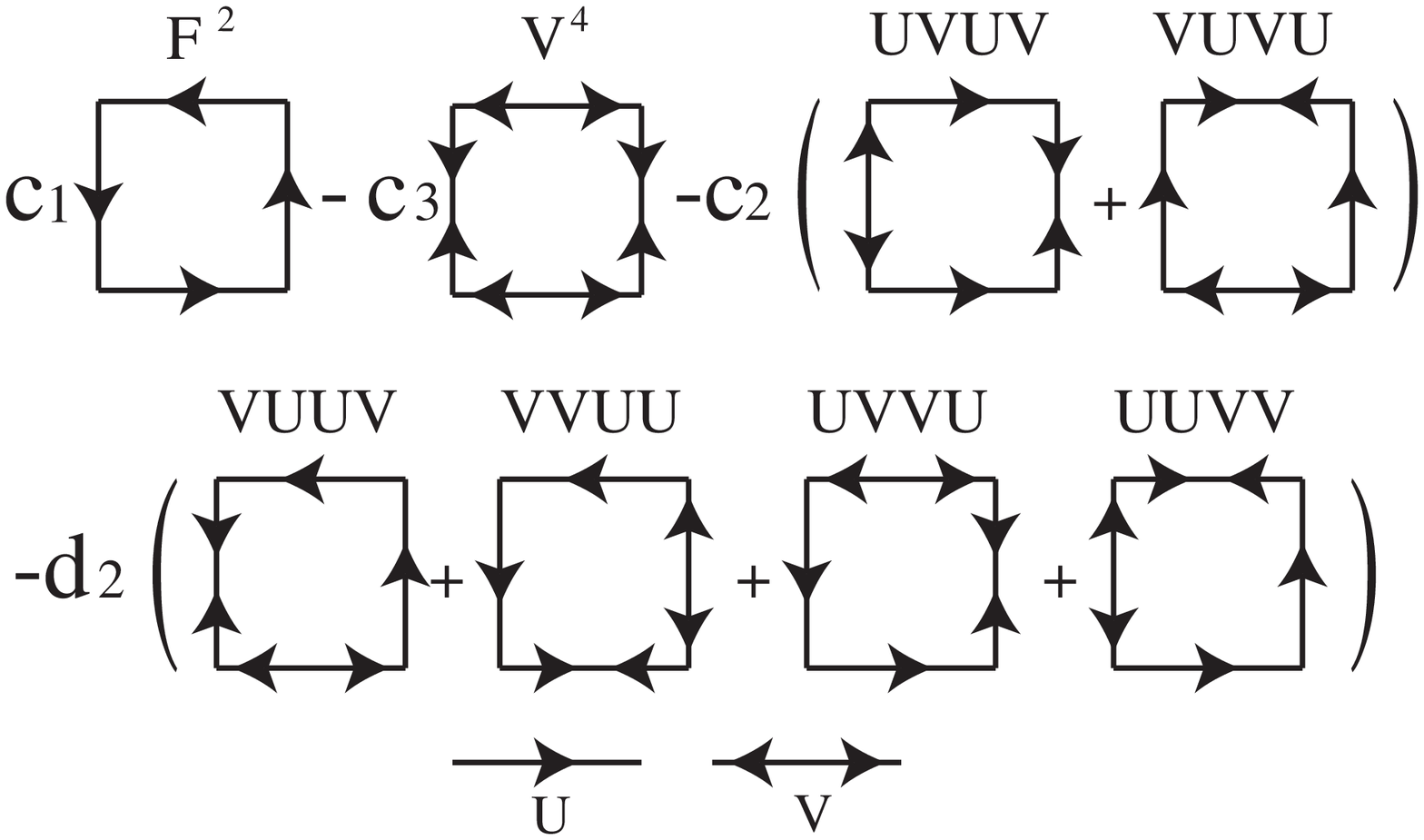}
\caption{$S_{\rm GL}$ in Eq.(\ref{action}).
}
\label{fig:SGL}
\end{center}
\end{figure}

In the previous paper\cite{UV3},
the following action of the GL theory of the UCSC was derived from a 
microscopic Hamiltonian by using the path-integral 
methods (see Fig.\ref{fig:SGL}),
\begin{eqnarray}
S_{\rm GL} &=&g\Big[ c_{1} \sum_{x,\mu \neq \nu}  F^2_{\mu\nu}(x) 
 - c_{2} \sum_{x,\mu \neq i} ( U_{x,\mu} V^{\ast}_{x+\mu,i}
U_{x+i,\mu} V_{x,i} + C.C. ) \nonumber \\
&& - d_{2} \sum_{x,i \neq j} ( U_{x,i} U_{x+i,j} 
V^{\ast}_{x+j,i} V_{x,j} + C.C. )  
 - c_{3} \sum_{x,i \neq j} ( V_{x,i} V^{\ast}_{x+i,j} 
V_{x+j,i} V^{\ast}_{x,j} + C.C. )\Big] \nonumber  \\
&& + \alpha \sum_{x,i} \left| V_{x,i} \right|^{4} - \beta \sum_{x,i} 
\left| V_{x,i} \right|^{2},
\label{action}
\end{eqnarray}
where $F_{\mu\nu}(x)=A_{x,\mu}+A_{x+\mu,\nu}-A_{x,\nu}-A_{x+\nu,\mu}$
($\mu,\nu=0,1,...,d$) and $U_{x,\mu}=e^{iA_{x,\mu}}$.
The following comments on $S_{\rm GL}$ are in order.
\begin{enumerate}
\item 
The $0$-th component vector potential $A_{x,0}$ mediates the 
long-range Coulomb interaction between the CP's.
\item
$S_{\rm GL}$ is invariant under the following gauge transformation,
\begin{equation}
A_{x,\mu}\rightarrow A_{x,\mu}+\alpha_{x+\mu}-\alpha_x,  \;\;\;
V_{x,i}\rightarrow e^{i\alpha_{x+i}}V_{x,i}e^{i\alpha_x}
\end{equation}
where $\alpha_x$ is an arbitrary function of $x$.
\item 
$c_1 \sim \beta$ in $S_{\rm GL}$ are parameters and the overall factor
$g$ plays a role of $1/\hbar$ and controls quantum 
fluctuations\cite{XY,UV3}.
\item
The $c_2$-term represents the hopping of $V_{x,i}$, whereas the $d_2$-term
determines the relative phase of the adjacent $V_{x,i}$'s, e.g. the negative
$d_2$ enhances $d$-wave condensation of $V_{x,i}$.
\item
The $c_3$-term in $S_{\rm GL}$ controls plaquette-flux 
(i.e., plaquette-vortex) of $V_{x,i}$.
\item
For large $c_3$, we expect that monopole configurations of $V_{x,i}$ are 
suppressed and effects of the compactness of $V_{x,i}$ becomes negligibly small.
\item
Parameters $\alpha$ and $\beta$ control the mean value of $V_{x,i}$
and its fluctuations.
\end{enumerate}

In the previous paper\cite{UV3}, we investigated the phase structure 
of the models
(\ref{action}) for $c_1=c_2=-d_2=c_3=1$ as varying the parameter $g$
and found existence of {\em second-order SC} quantum phase transitions.
In the $d=3$ cases, however, we found another phase transition within the
SC phase besides the SC transition and we discussed its physical meanings.
Though the high-$T_c$ cuprates are quasi-two-dimensional and have
layered structure, critical behavior is governed by its three dimensionality.
Therefore the above additional phase transition might be observed in
experiments.

In this paper we shall continue the previous study and investigate phase 
structure and quantum critical behavior of the model for various parameters 
regions.
In particular, we shall discuss and verify that the locations of the 
two second-order phase transitions get closer with each other and finally 
they merge to a single  first-order phase transition as the value of $c_3$ 
is increased.

This paper is organized as follows.
In Sec.2, we shall study the London limit (the CP field $|V_{x,i}|=1$) of
the model in $(3+1)$D.
For the case $c_1=c_2=-d_2=c_3=1$, there are two second-order phase transitions
as observed in the case investigated in the previous paper.
We discuss the physical meanings of the phase transitions.
Then we study how the phase structure will change as the value $c_3$ is 
increased.
In Sec.3, we shall study the case $\alpha=\beta=5$ and show that
various phase structures appear as the value of $c_3$ is varied.
There new type of phase transitions appear, which are relevant for 
SC materials with strong three-dimensionality and to be observed 
by experiments.
Section 4 is devoted for conclusion.

\section{London limit in (3+1)D}
\setcounter{equation}{0}
In the previous paper\cite{UV3}, we reported the results of study on 
the $\alpha=\beta=5$ and $\alpha=\beta=10$ cases.
We investigated the case with the parameters $c_1=c_2=-d_2=c_3=1$,
and found that in $(3+1)$D there are two phase transitions at 
$g_c$ and $g_c'$ ($g_c < g_c'$). 
By the calculation of the gauge-boson mass $M_{\rm G}$, we concluded that
the first transition at $g_c$ is the genuine SC phase transition,
which is connected with the proliferation of {\em vortices} of the 
CP field $V_{x,i}$,
whereas the second one at $g_c'$ is related to the {\em monopoles} 
proliferation of $V_{x,i}$.
Furthermore we inferred that the locations of the two phase transitions would  
be getting closer and finally they would merge into a single {\em first-order}
phase transition for larger values of $c_3>1$.
Recently this kind of phenomena of the phase transition were observed
in simpler gauge-theory systems\cite{Higgs2}.

In this section we shall study the system $S_{\rm GL}$ in the London limit
$|V_{x,i}|=1$.
Then the system given by Eq.(\ref{action}) can be regarded as a 
U(1)$_{\rm NC}\times$U(1)$_C$ gauge model where the EM vector potential
$A_{x,\mu}$ corresponds to the $(3+1)$D noncompact U(1) gauge field, whereas
the CP field $V_{x,i}$ is the 3D compact U(1) gauge field.
We shall study the system and determine its phase structure
by calculating ``internal energy" $E$, ``specific heat" $C$,
the magnetic penetration depth, etc.
$E$ and $C$ are defined as follows,
\begin{equation}
E \equiv -\langle S_{\rm GL} \rangle/L^3, \;\;\;
C \equiv \langle (S_{\rm GL}-\langle S_{\rm GL}\rangle)^2\rangle/L^3,
\label{EC}
\end{equation}
where $L^3$ is the system size.
The gauge-boson mass $M_{\rm G}$ is nothing but the inverse magnetic
penetration depth and is obtained by calculating the correlation function 
of the magnetic field $F_{ij}(x)$\cite{MG}.
Other interesting quantities are monopole density of $U_{x,i}$
and $V_{x,i}$, $\rho_U$ and $\rho_V$, which measure the strength of the 
topologically nontrivial fluctuations of the link gauge 
fields\cite{monopole,UV1}.
$\rho_U$ must have strong correlation to $M_{\rm G}$ and have
small values, but sometimes
$\rho_U$ gives clearer signal of the phase transition (see later discussion).
As the CP $V_{x,i}$ is defined on lattice link in the present system, 
vortices (more precisely vortex lines) in the SC phase can be terminate at 
(anti)monopoles.
Therefore $\rho_V$ is a very specific quantity in the present model 
$S_{\rm GL}$ and does not exists in other systems like the XY model.
However as $c_3$ is getting large, $V_{x,i}$ tends to have ``{\em pure-gauge}
configurations" like $V_{x,i}\sim e^{i\theta_{x+i}}e^{i\theta_x}$\cite{PG} and 
$\rho_V$ decreases very rapidly.

\begin{figure}[htbp]
\begin{center}
\epsfxsize=6cm
\epsffile{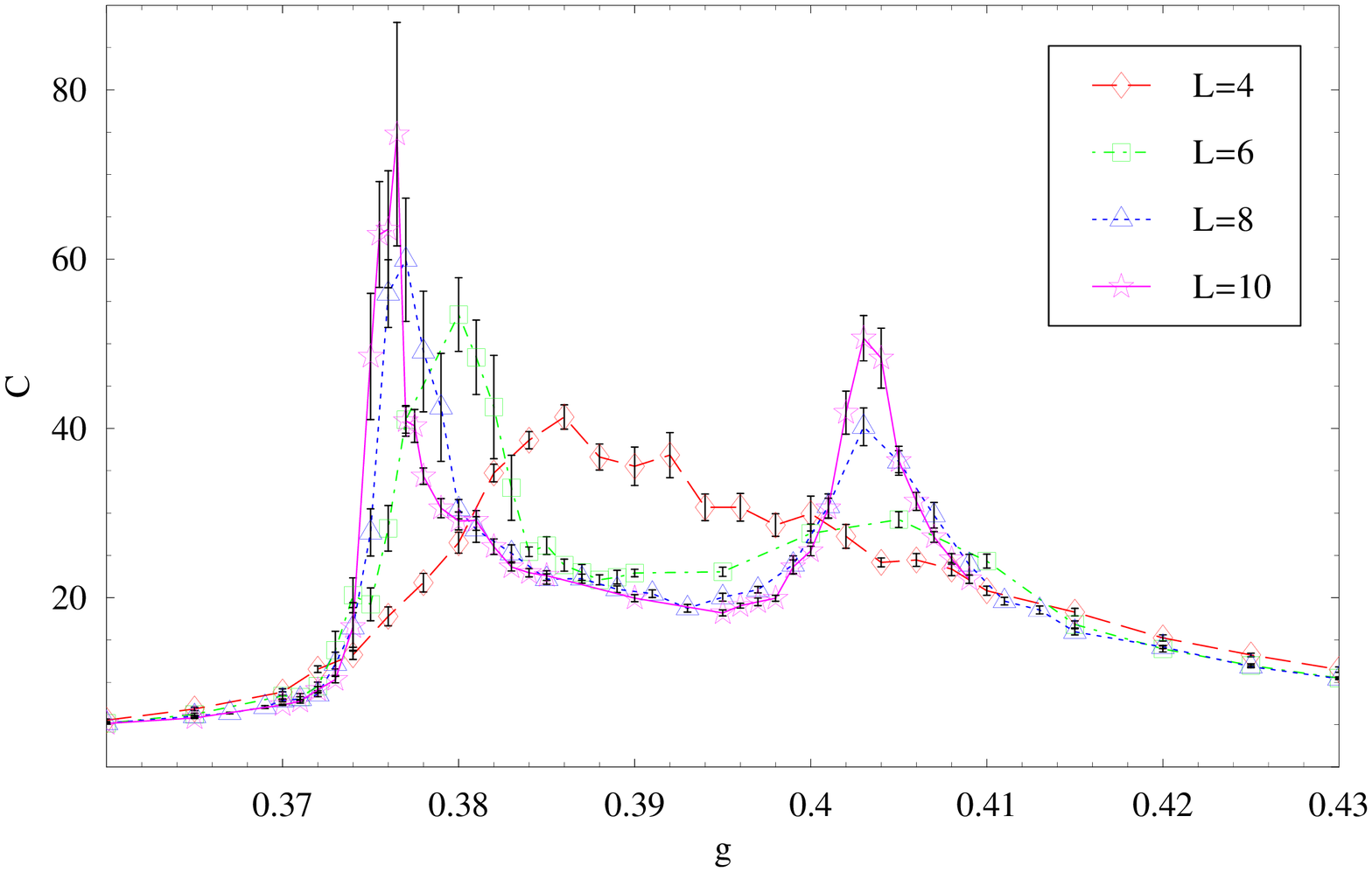}
\epsfxsize=6cm
\epsffile{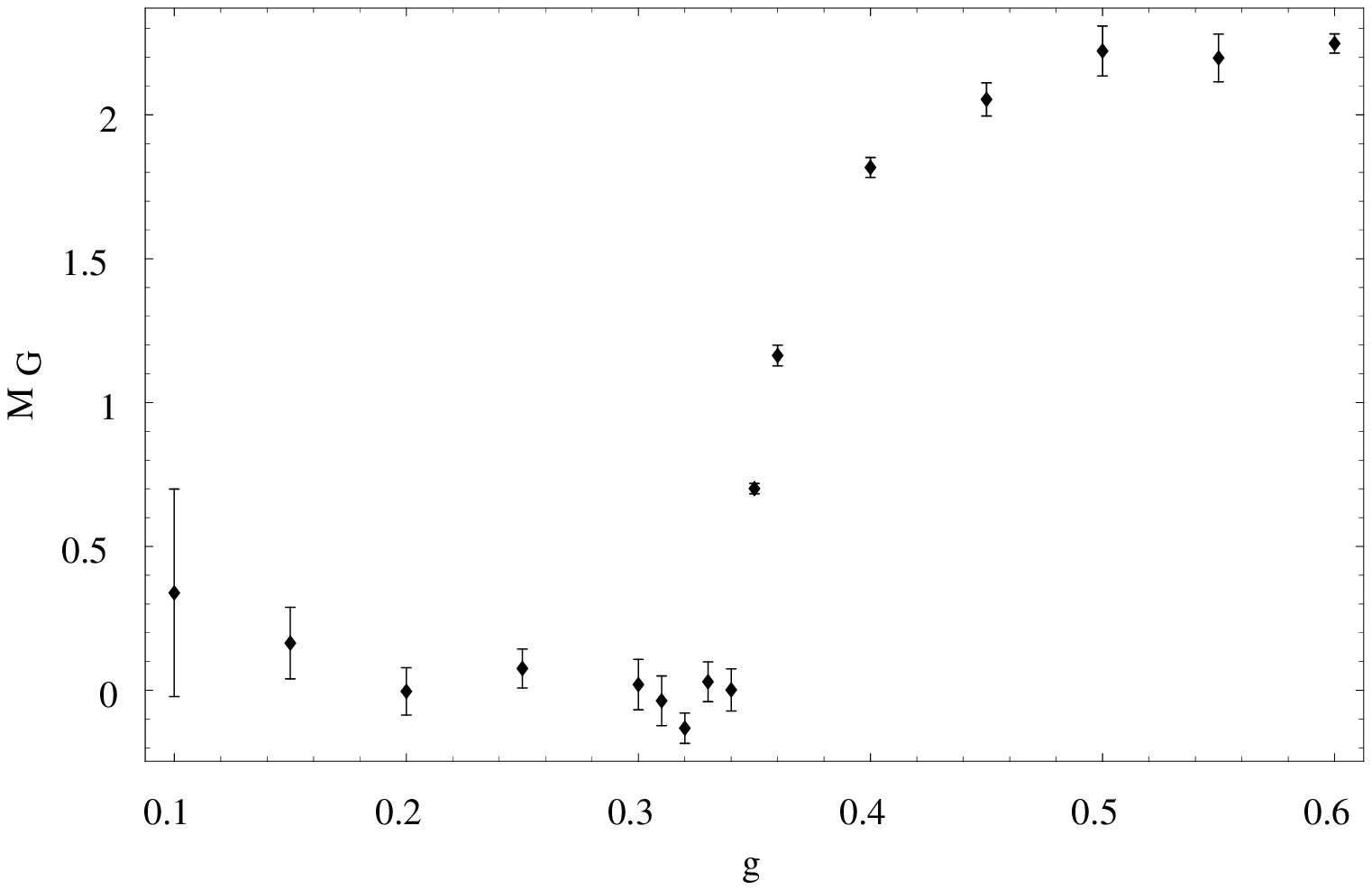}
\caption{(color online)
$C$ for the London limit with $c_3=1$.
There are two peaks both of which have system-size dependence (left).
Gauge-boson mass $M_{\rm G}$. It develops nonvanishing value 
at the first critical coupling $g_c \sim 0.37$ (right).
}
\label{fig:London1}
\end{center}
\end{figure}

\begin{figure}[htbp]
\begin{center}
\epsfxsize=6cm
\epsffile{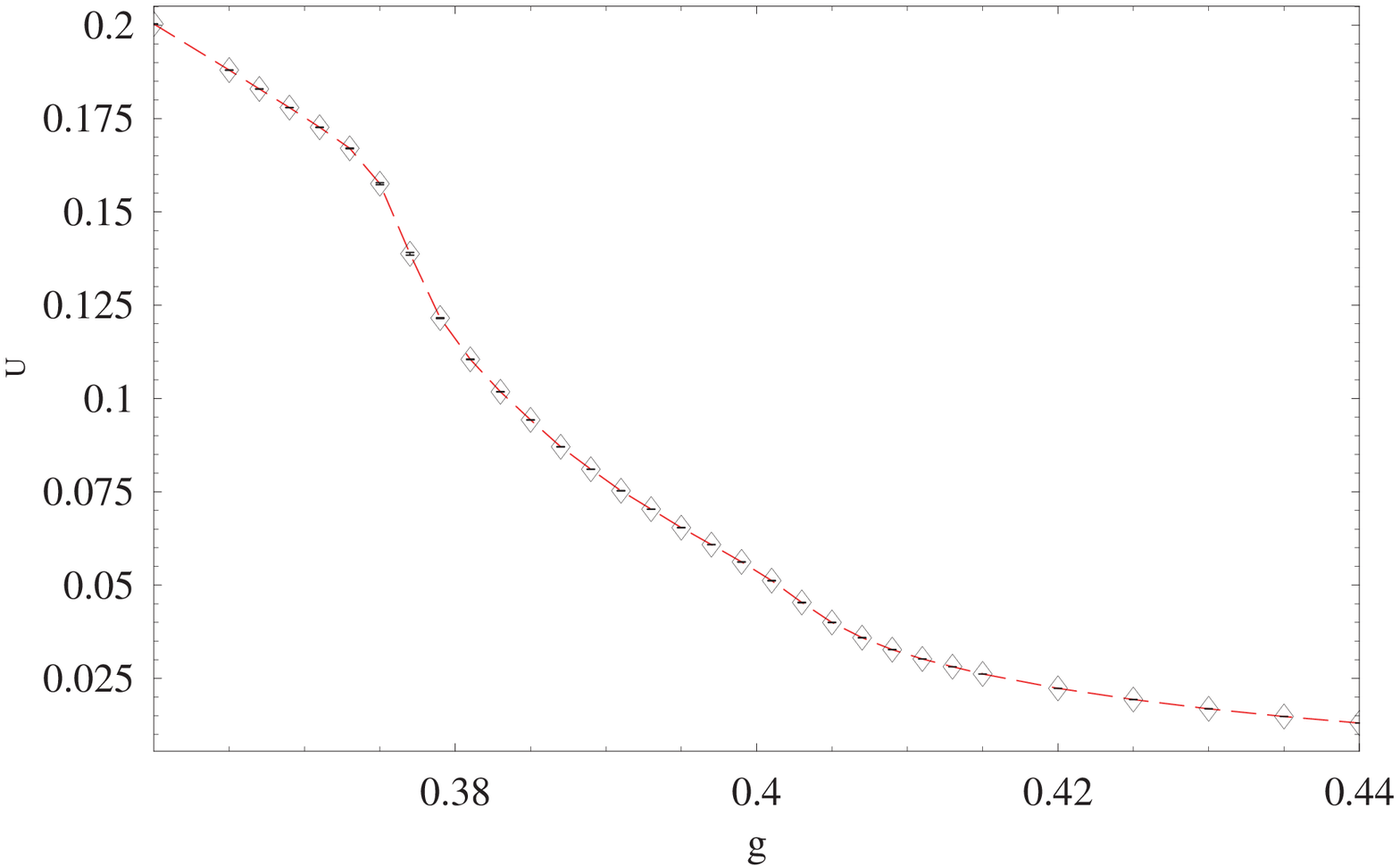}
\epsfxsize=6cm
\epsffile{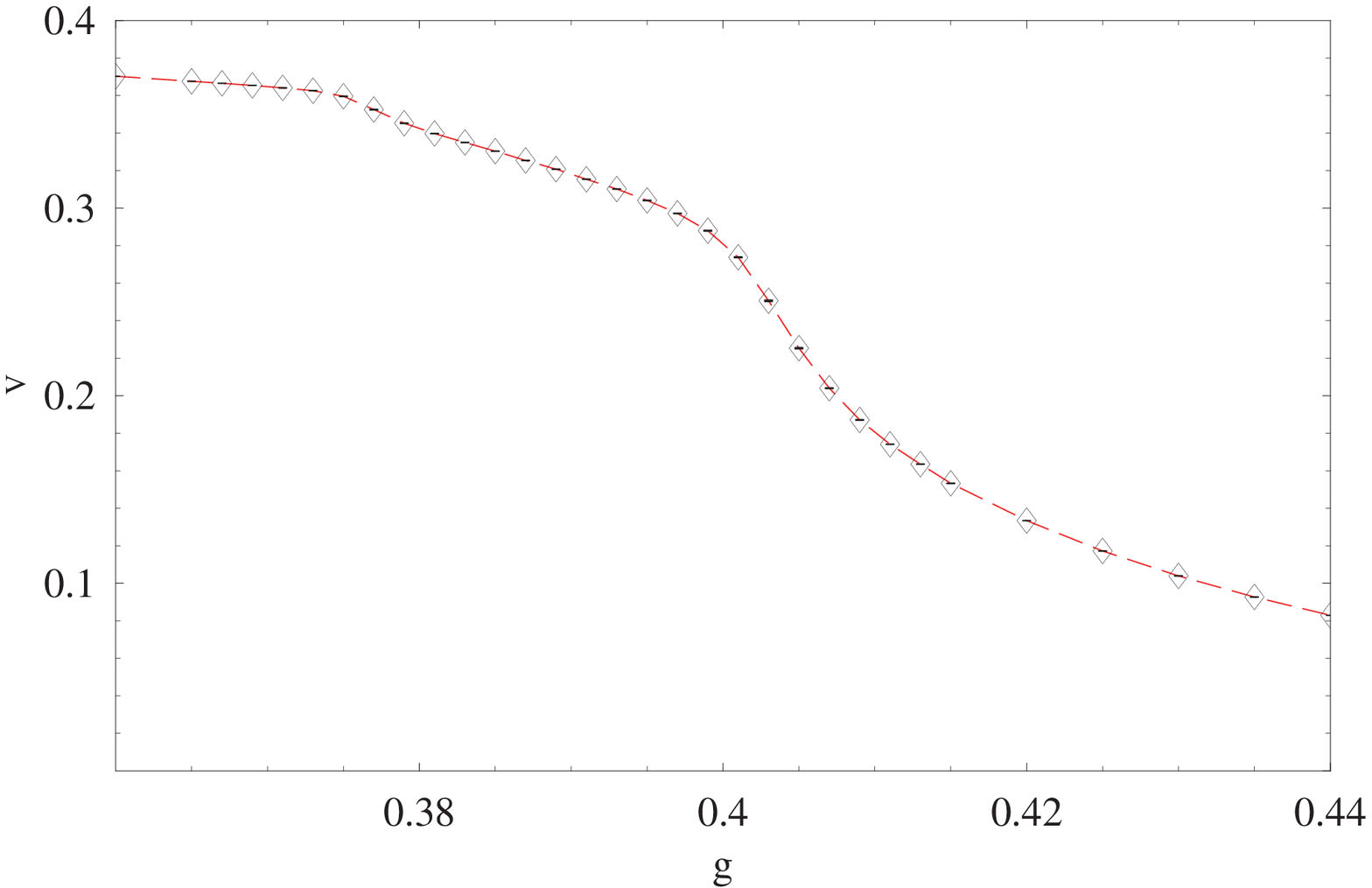}
\caption{Monopole densities in the London limit with $c_3=1$.
$\rho_U$ (left) and $\rho_V$ (right).
In the region between $g_c=0.37$ and $g_c'=0.405$, $\rho_V$
has finite values.
}
\label{fig:LondonIns}
\end{center}
\end{figure}

We first study the case $c_1=c_2=-d_2=c_3=1$.
The calculations of $C$ and the inverse penetration depth $M_{\rm G}$
are shown in Fig.\ref{fig:London1}.
From the result of $C$,
it is obvious that there are two second-order phase transitions
at $g_c\sim 0.37$ and $g_c'\sim 0.405$, since the each peak develops
as the system size $L$ is getting large.
From the calculation of $M_{\rm G}$, we can see that the
first transition at $g_c$ is the genuine SC transition.
In order to see the physical meaning of the second transition at 
$g_c'$, we measured the $U$ and $V$ monopole densities
$\rho_U, \; \rho_V$.
See Fig.\ref{fig:LondonIns}.
$\rho_U$ has small values for $U_{x,\mu}$ is the noncompact gauge field
and it changes the behavior at both $g_c$ and $g_c'$.
$\rho_V$ has finite (fairly large) values in the region between $g_c$ and
$g_c'$ and it starts to decrease at $g_c'$.
This indicates that the transition at $g_c'$ corresponds to the
monopole proliferation-suppression of the CP field $V_{x,i}$, i.e., 
though the density of vortices decreases at the 
first phase transition point $g_c$, short vortex lines, which terminate at 
(anti)monopoles, survive in the region between $g_c$ and $g_c'$.

Let us turn to the case $c_3>1$.
As explained in the above, effects of the compactness of the
``gauge field" $V_{x,i}$ is suppressed for large $c_3$, and the pure-gauge
configurations $V_{x,i}\sim e^{i\theta_{x+i}}e^{i\theta_x}$ dominate
the path integral.
Then one may naively expect that the system is getting close to the 4D XY 
model.
On the other hand, the dynamics of $V_{x,i}$ is getting close to
that of the noncompact U(1) gauge field like $U_{x,i}$ because of the
suppression of the $V$-monopoles, 
and as a result the system becomes $U-V$ symmetric. 
Then one may expect that two second-order 
phase transitions in the $c_3=1$ case tend to get close with each other and 
finally they merge into a single phase transition.
From the above point of view, it is very interesting to investigate the 
cases of larger value of $c_3$.

\begin{figure}[htbp]
\begin{center}
\epsfxsize=6cm
\epsffile{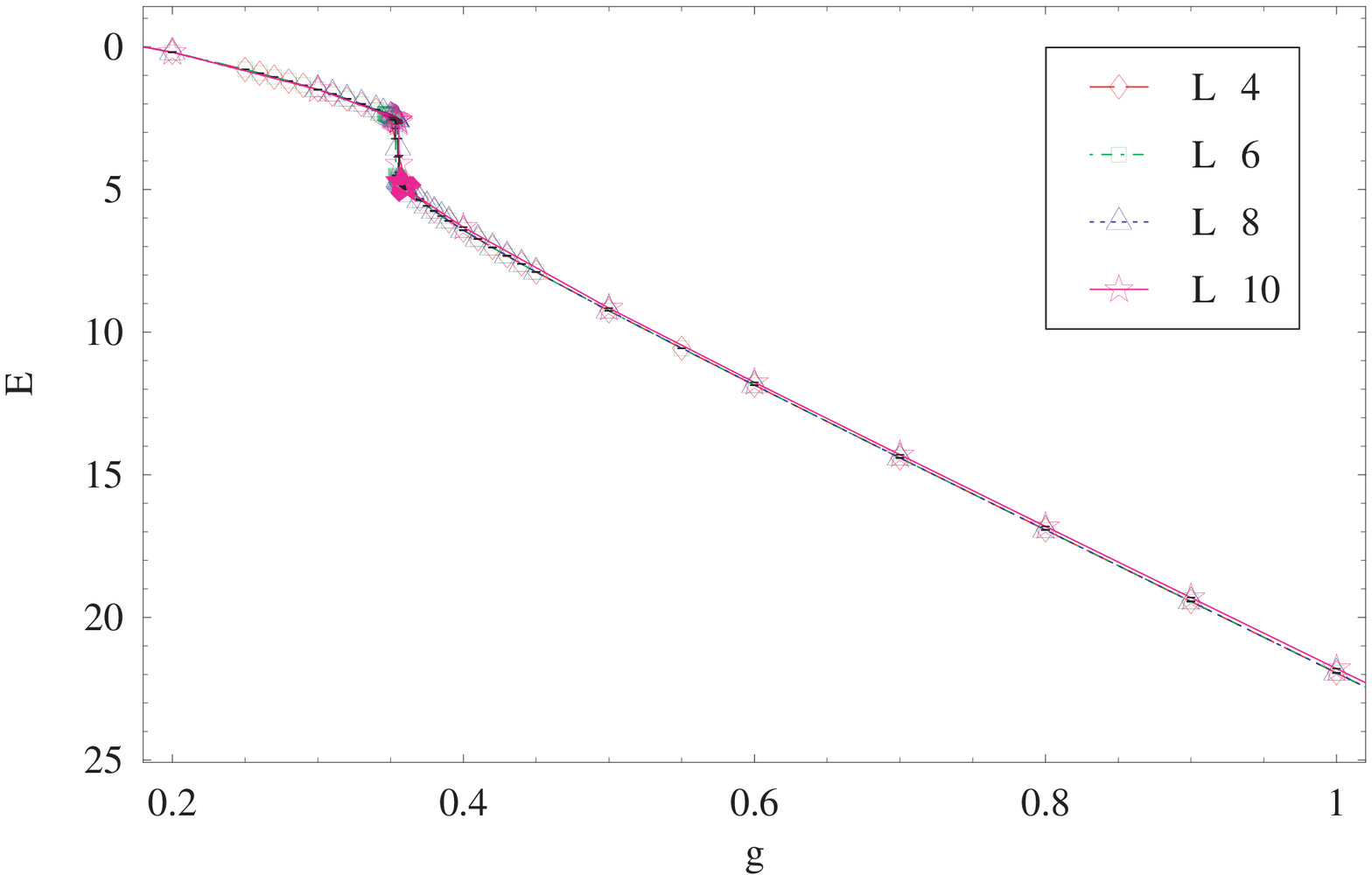}
\epsfxsize=6cm
\epsffile{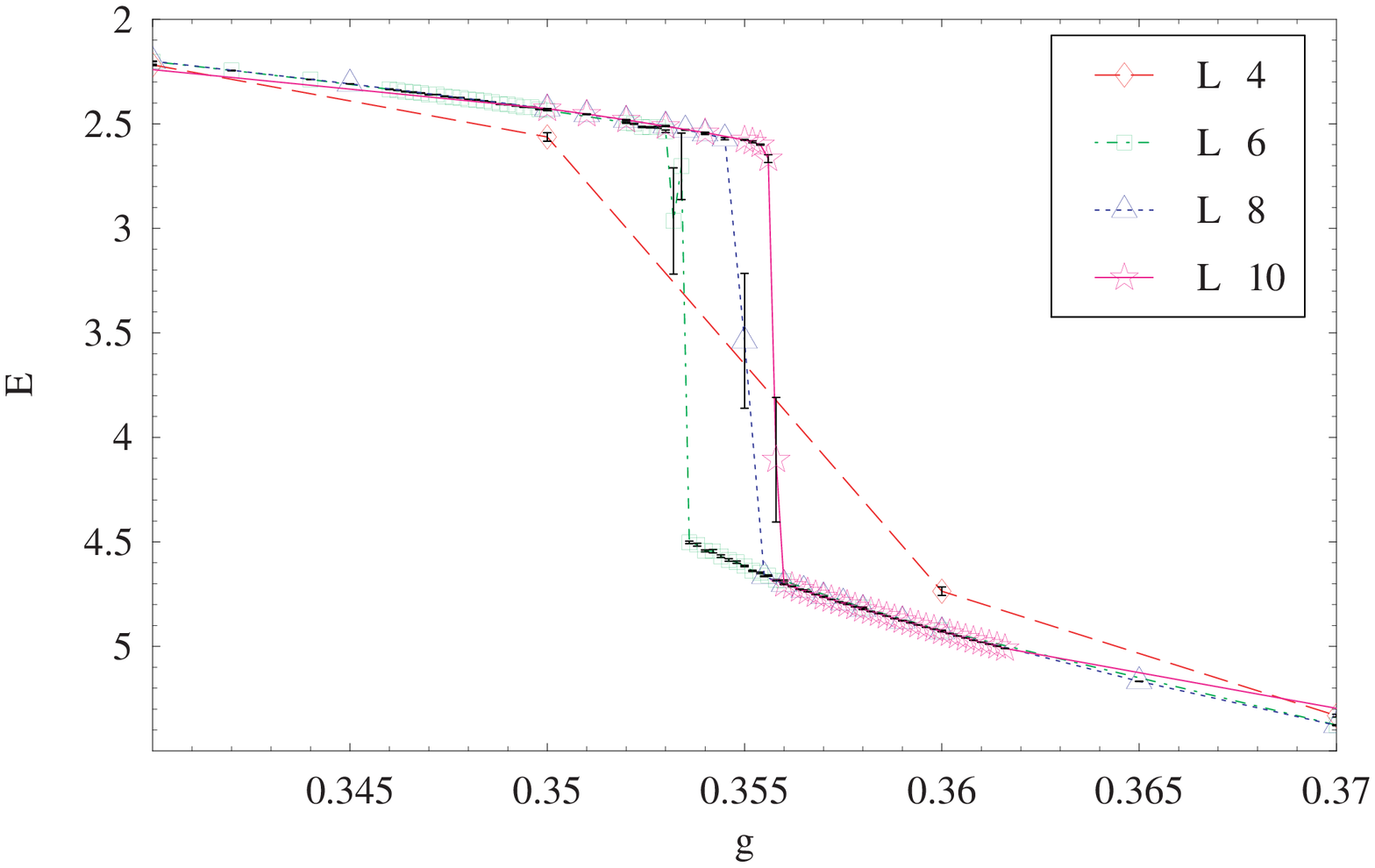}
\caption{(color online)
Internal energy $E$ for $c_3=2$ in the London limit.
As the system size is getting larger, the discontinuity at 
$g_c=0.355$ becomes sharper.
}
\label{fig:London2E}
\end{center}
\end{figure}

We first show the results of the $c_3=2$ case.
Internal energy $E$ is shown in Fig.\ref{fig:London2E}.
These results show that there exists a finite jump in $E$ at $g=0.355$
and its discontinuity gets sharper as the system size $L$ is increased.
This means that there exists a first-order phase transition at $g_c=0.355$.
No other anomalous behavior is observed in $E$ and $C$ as $g$ is increased
further.

\begin{figure}[htbp]
\begin{center}
\epsfxsize=6cm
\epsffile{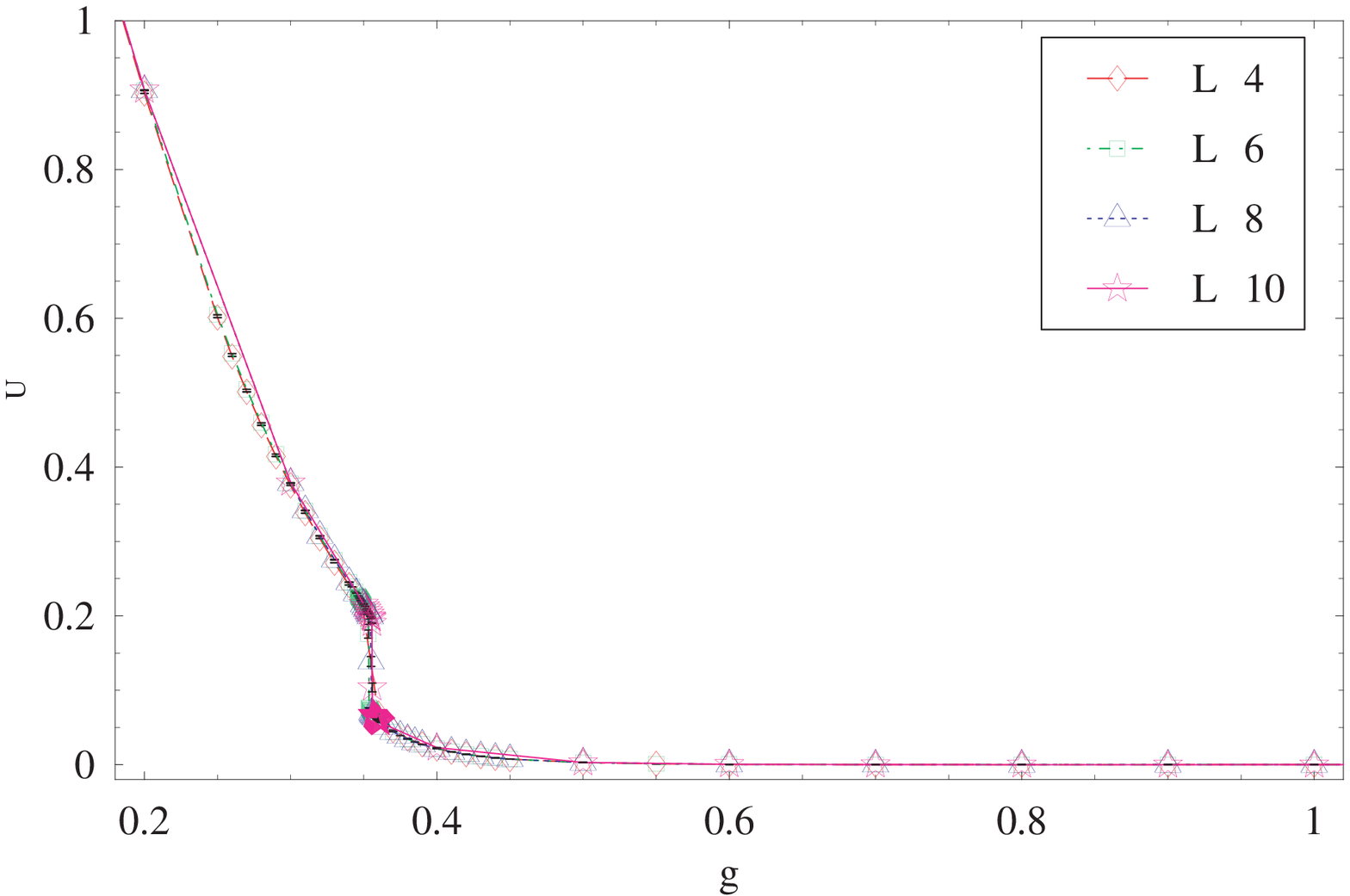}
\epsfxsize=6cm
\epsffile{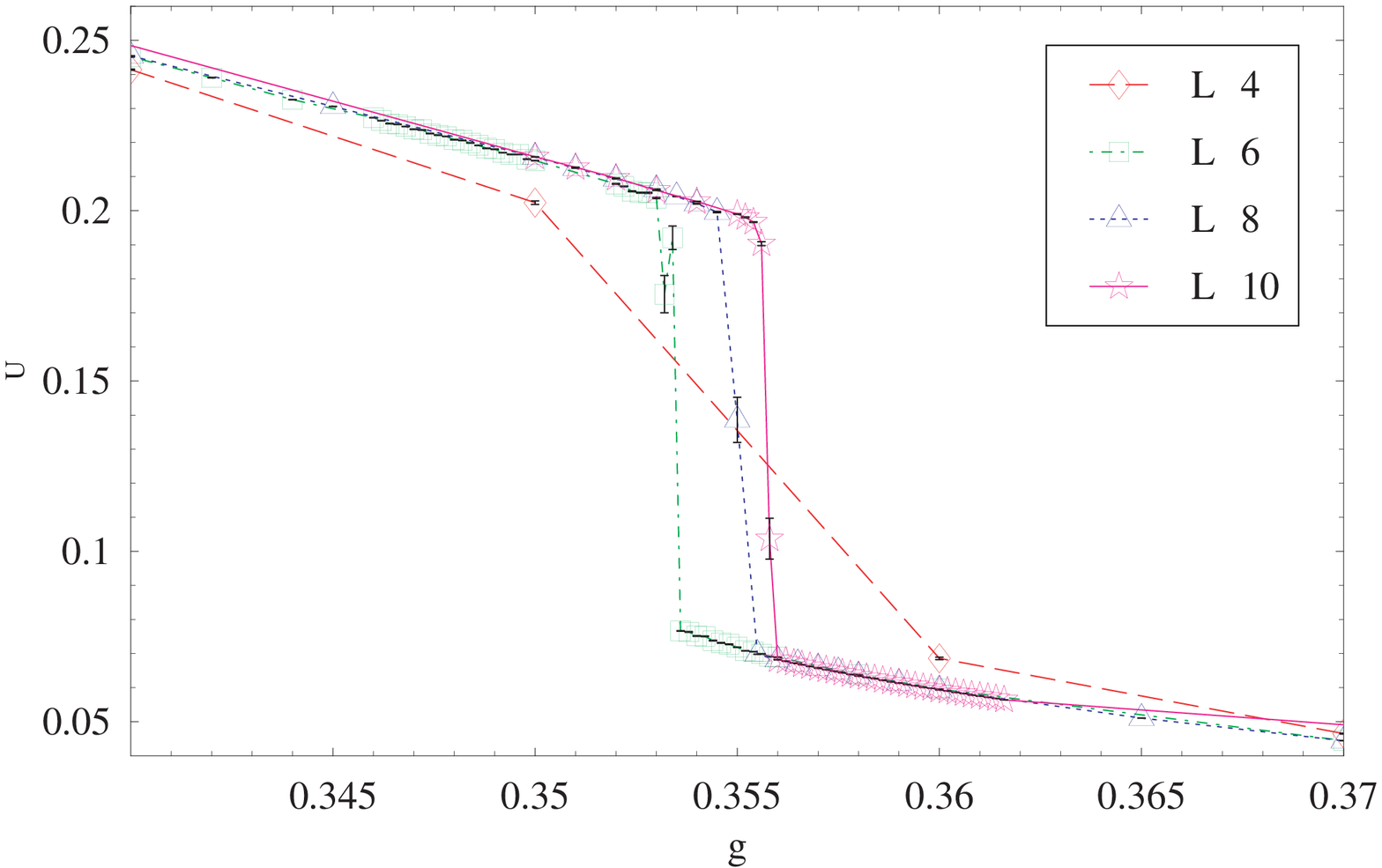}
\caption{(color online)
The U-monopole density $\rho_U$ for $c_3=2$ in the London limit.
As the system size is getting larger, the discontinuity at 
$g_c=0.355$ becomes sharper.
}
\label{fig:London2IU}
\end{center}
\end{figure}

\begin{figure}[htbp]
\begin{center}
\epsfxsize=6cm
\epsffile{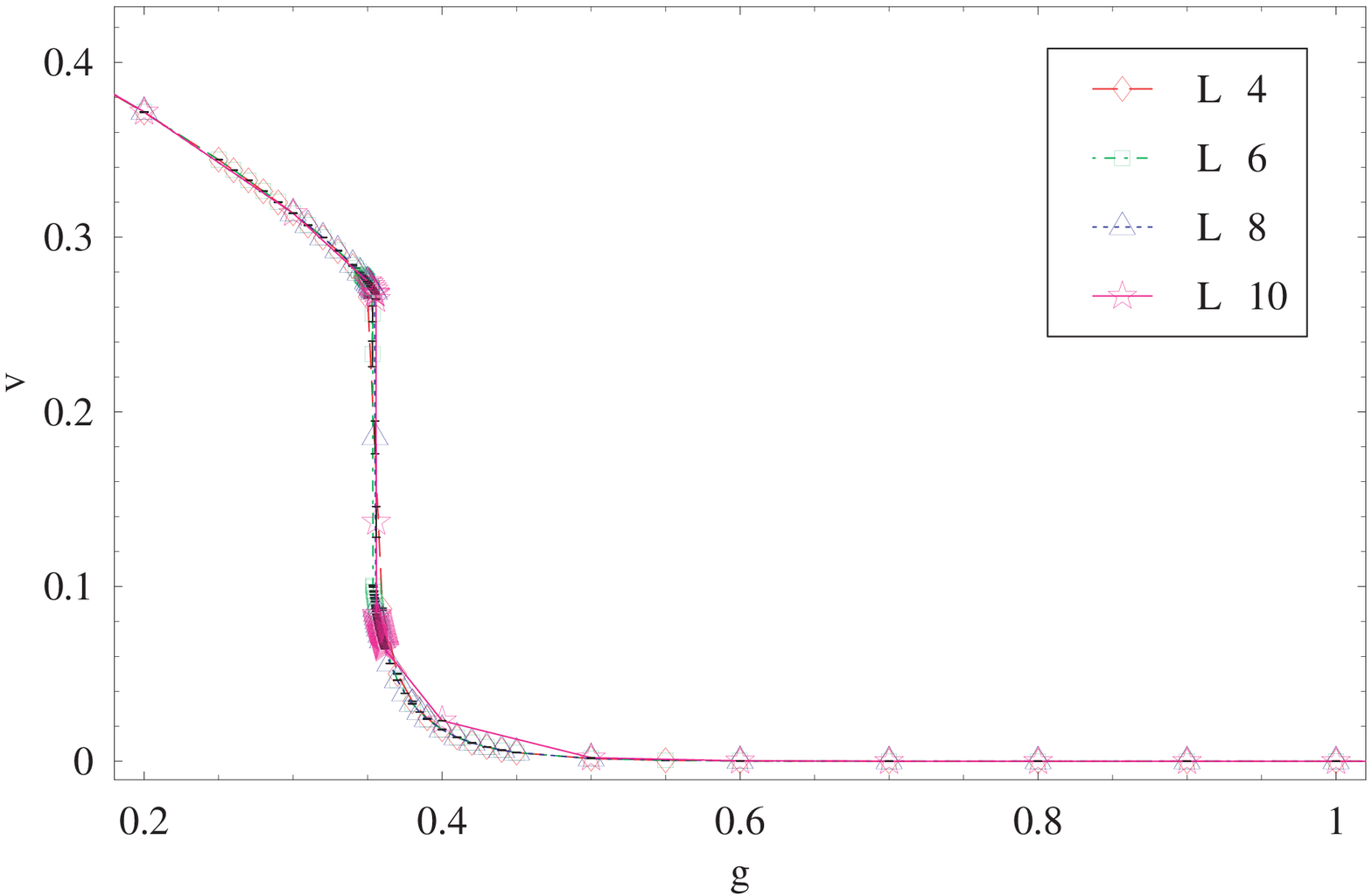}
\epsfxsize=6cm
\epsffile{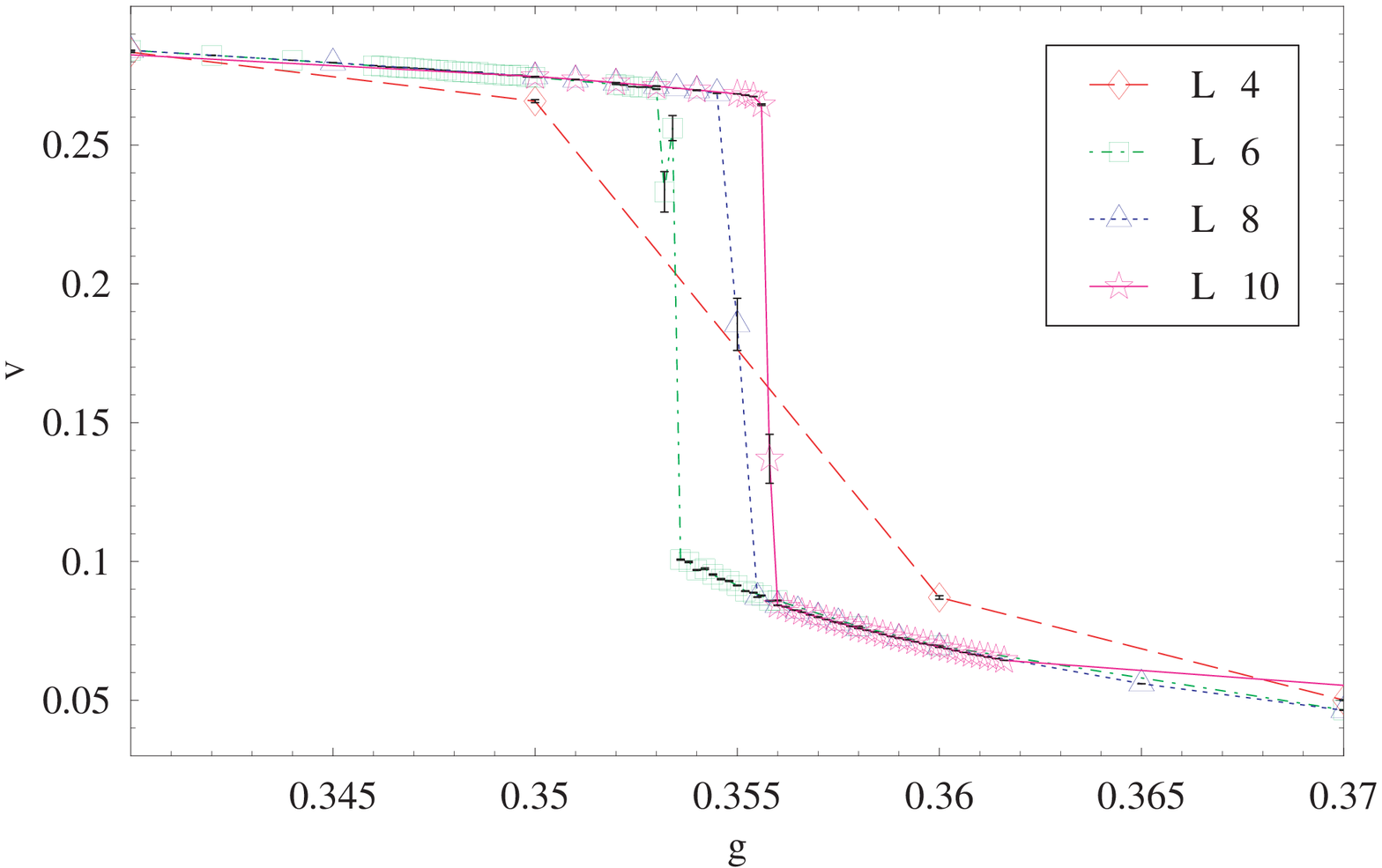}
\caption{(color online)
The V-monopole density $\rho_V$ for $c_3=2$ in the London limit.
As the system size is getting larger, the discontinuity at 
$g_c=0.355$ becomes sharper.
}
\label{fig:London2IV}
\end{center}
\end{figure}

\begin{figure}[htbp]
\begin{center}
\epsfxsize=7cm
\epsffile{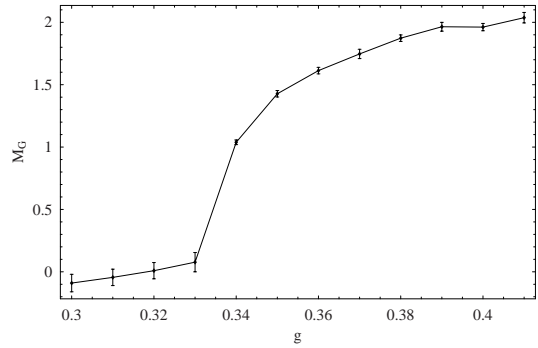}
\caption{The gauge-boson mass $M_{\rm G}$ for $c_3=2$
in the London limit.
}
\label{fig:London2M}
\end{center}
\end{figure}

We expect that two second-order phase transitions, which exist for $c_3=1$, 
merge into the first-order phase transition in the present $c_3=2$ case.
In order to verify the above expectation, we measured the monopole
densities $\rho_U$ and $\rho_V$.
See Figs.\ref{fig:London2IU} and \ref{fig:London2IV}.
It is obvious that both $\rho_U$ and $\rho_V$ exhibit
discontinuity at the phase transition point $g_c=0.355$ and are
vanishingly small for $g>g_c$.
We also measured the gauge-boson mass $M_{\rm G}$.
The result is shown in Fig.\ref{fig:London2M}.
$M_{\rm G}$ does not show sharp discontinuity at the critical point,
but it starts to increase from zero at $g_c=0.355$.


\begin{figure}[htbp]
\begin{center}
\epsfxsize=6cm
\epsffile{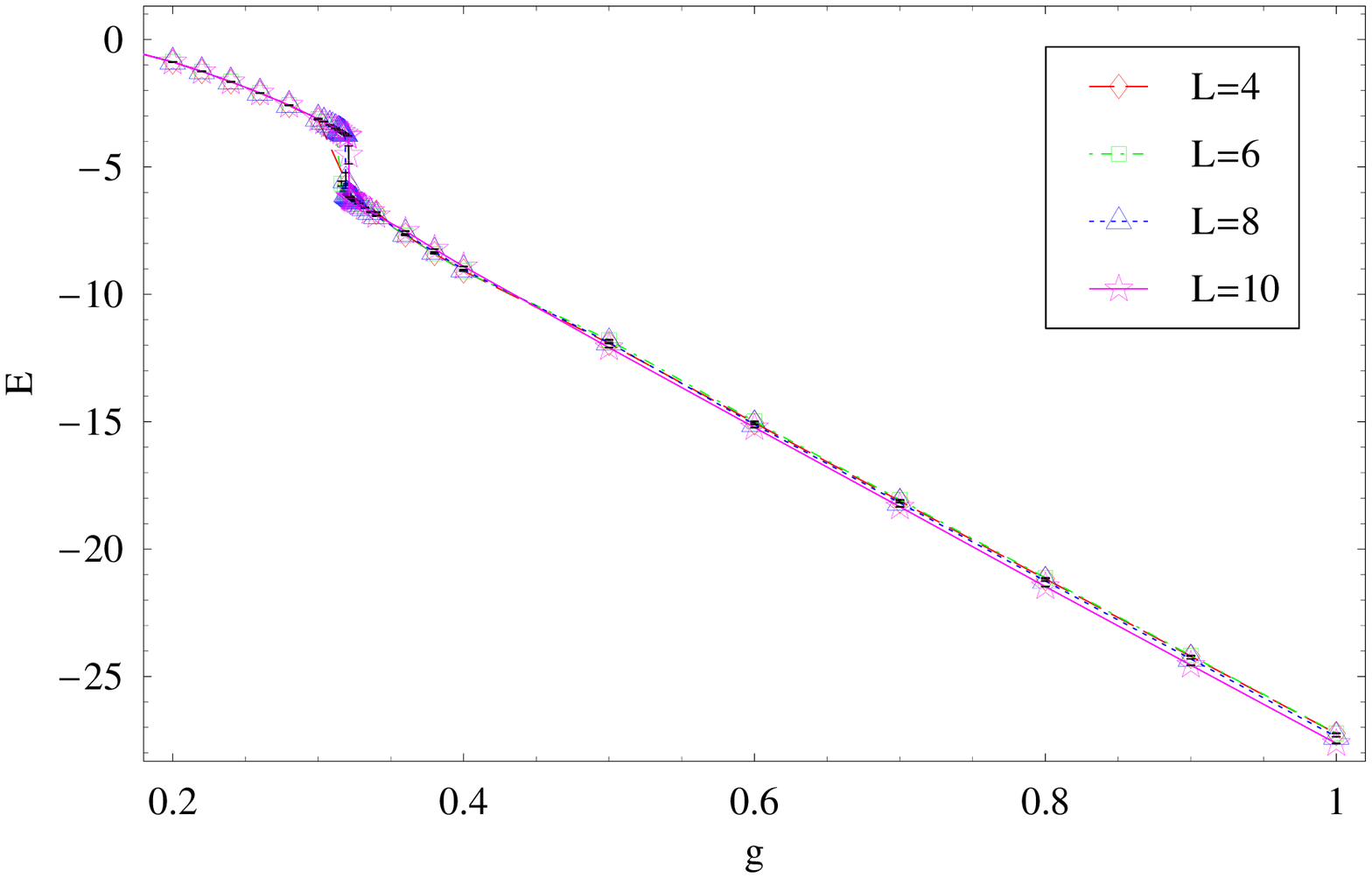}
\epsfxsize=6cm
\epsffile{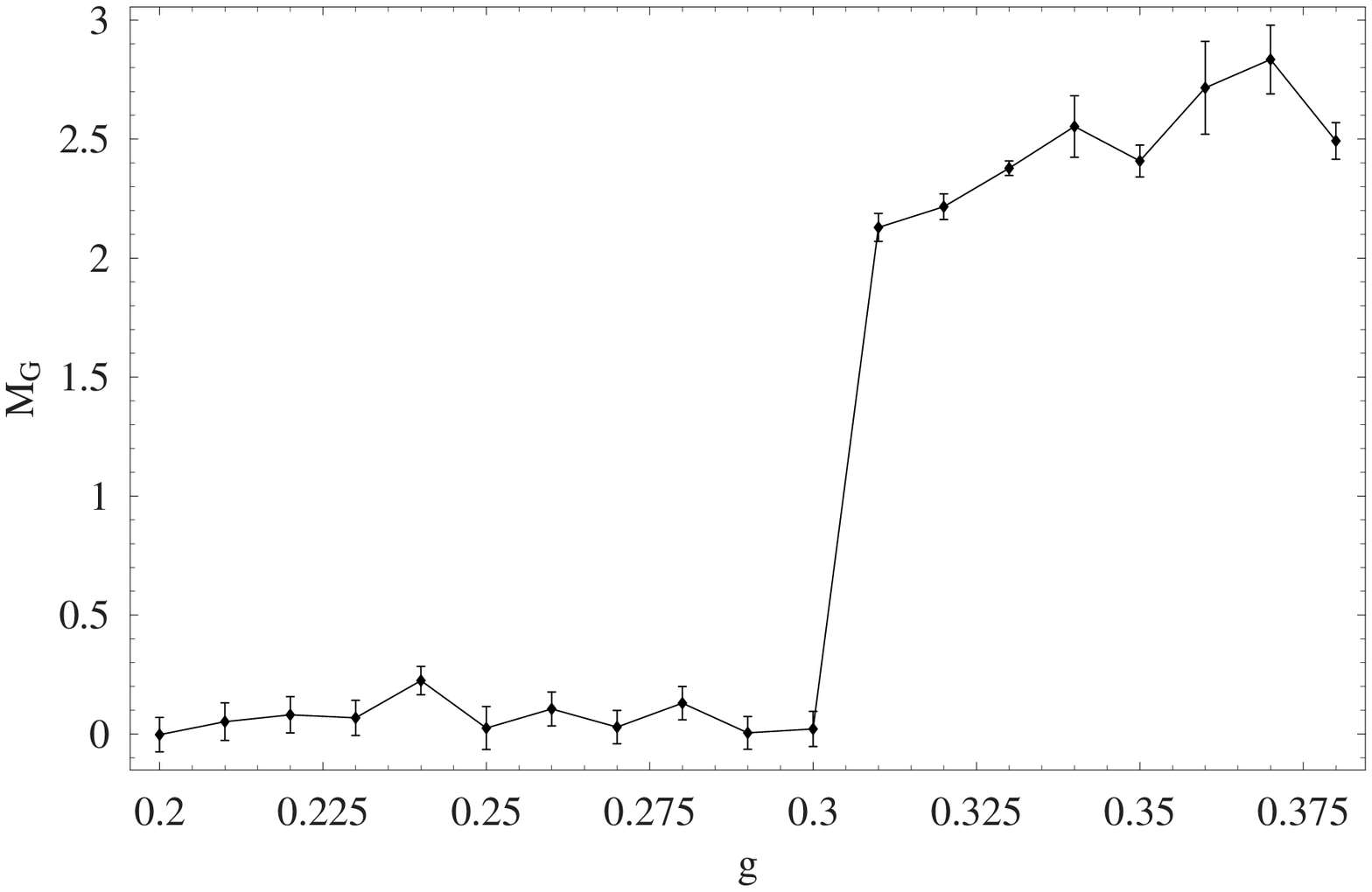}
\caption{(color online)
Internal energy $E$ and gauge-boson mass $M_{\rm G}$ for $c_3=4$
in the London limit.
Both quantities exhibit sharp discontinuity at critical point 
$g_c=0.320$.
}
\label{fig:London4E}
\end{center}
\end{figure}

We have also investigated the case of the London limit with $c_3=4$.
We show the results in Fig.\ref{fig:London4E}.
There is a first-order phase transition at $g=0.320$ and 
behavior of $E, \; \rho_U$ and $\rho_V$ have similar behavior
with those of the $c_3=2$.
However $M_{\rm G}$ shows a sharp discontinuity at the critical point
in the present case.

\section{Phase structure of $\alpha=\beta=5$ case} 
\setcounter{equation}{0}

In the previous section, we showed that the two second-order phase
transitions merge into a single first-order transition as $c_3$
is increased.
This behavior is expected from the recent studies on more tractable
gauge-Higgs models.
In this section, we shall study the system with 
$\alpha=\beta=5$ and see how the phase structure changes as 
the parameter $c_3$ is varied.
In this case $V_{x,i}$ is parameterized as  
$V_{x,i}=r_{x,i}e^{i\varphi_{x,i}}$ and the amplitude $r_{x,i}$
is also a dynamical variable and fluctuates around its mean value.
This may change the critical behavior of the model.
In fact, it is known that the phase structure of the 3D U(1) Higgs model 
depends on the potential term of $r_{x,i}$\cite{pHiggs}.

\begin{figure}[htbp]
\begin{center}
\epsfxsize=6cm
\epsffile{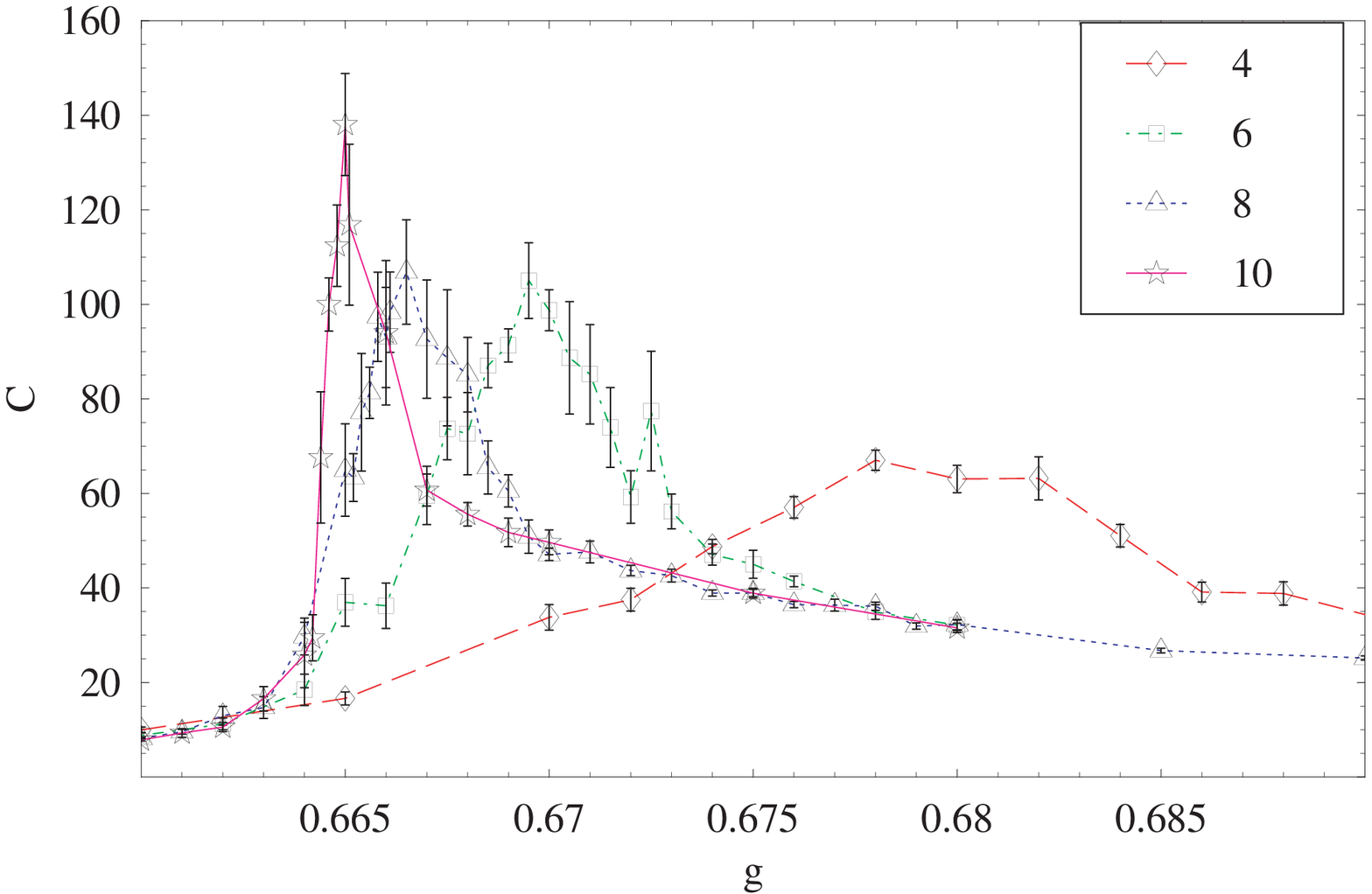}
\epsfxsize=6cm
\epsffile{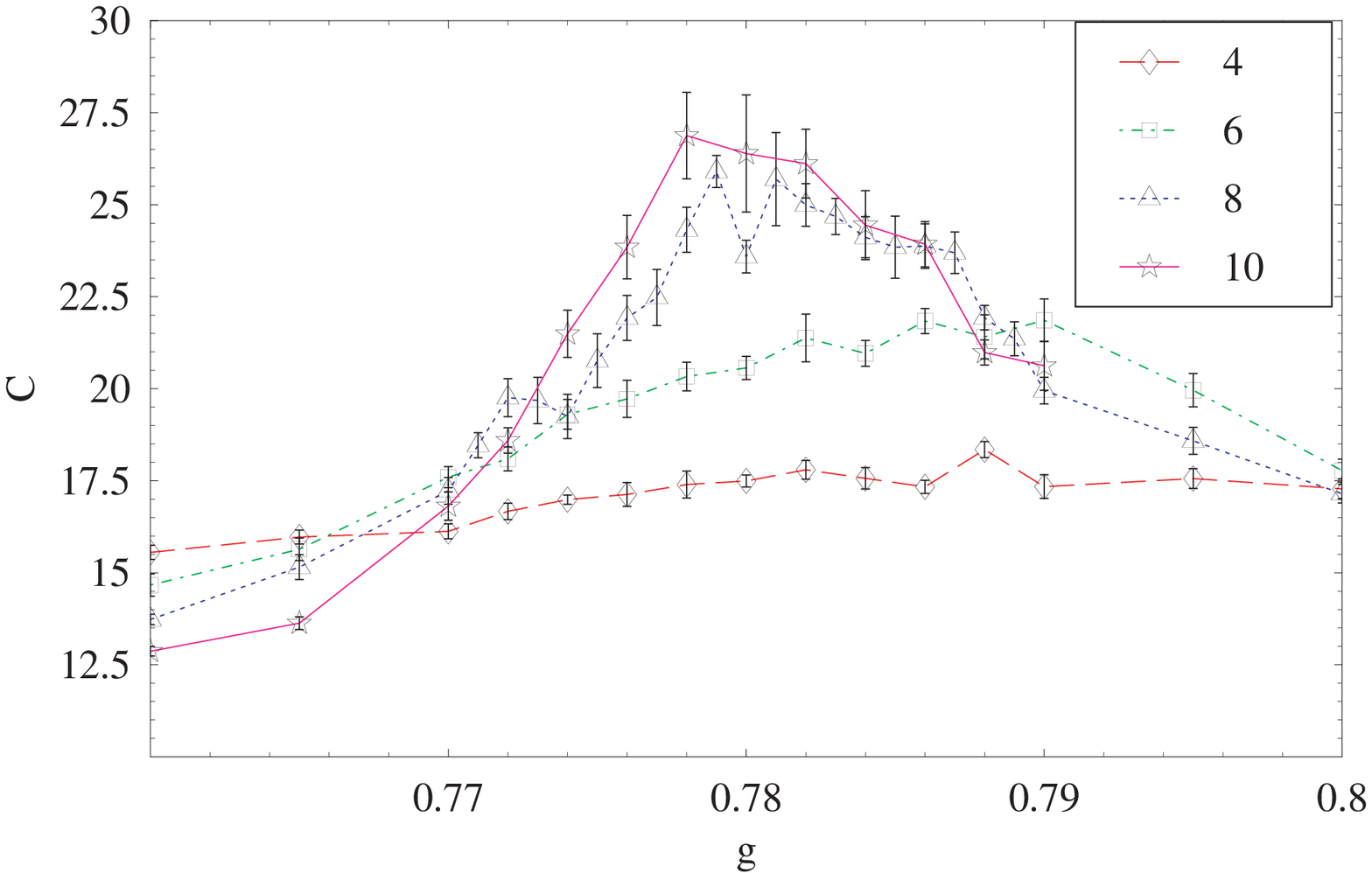}
\caption{(color online)
Two second-order phase transitions for $c_3=1, \alpha=\beta=5$.
}
\label{fig:28}
\end{center}
\end{figure}

\begin{figure}[htbp]
\begin{center}
\epsfxsize=6cm
\epsffile{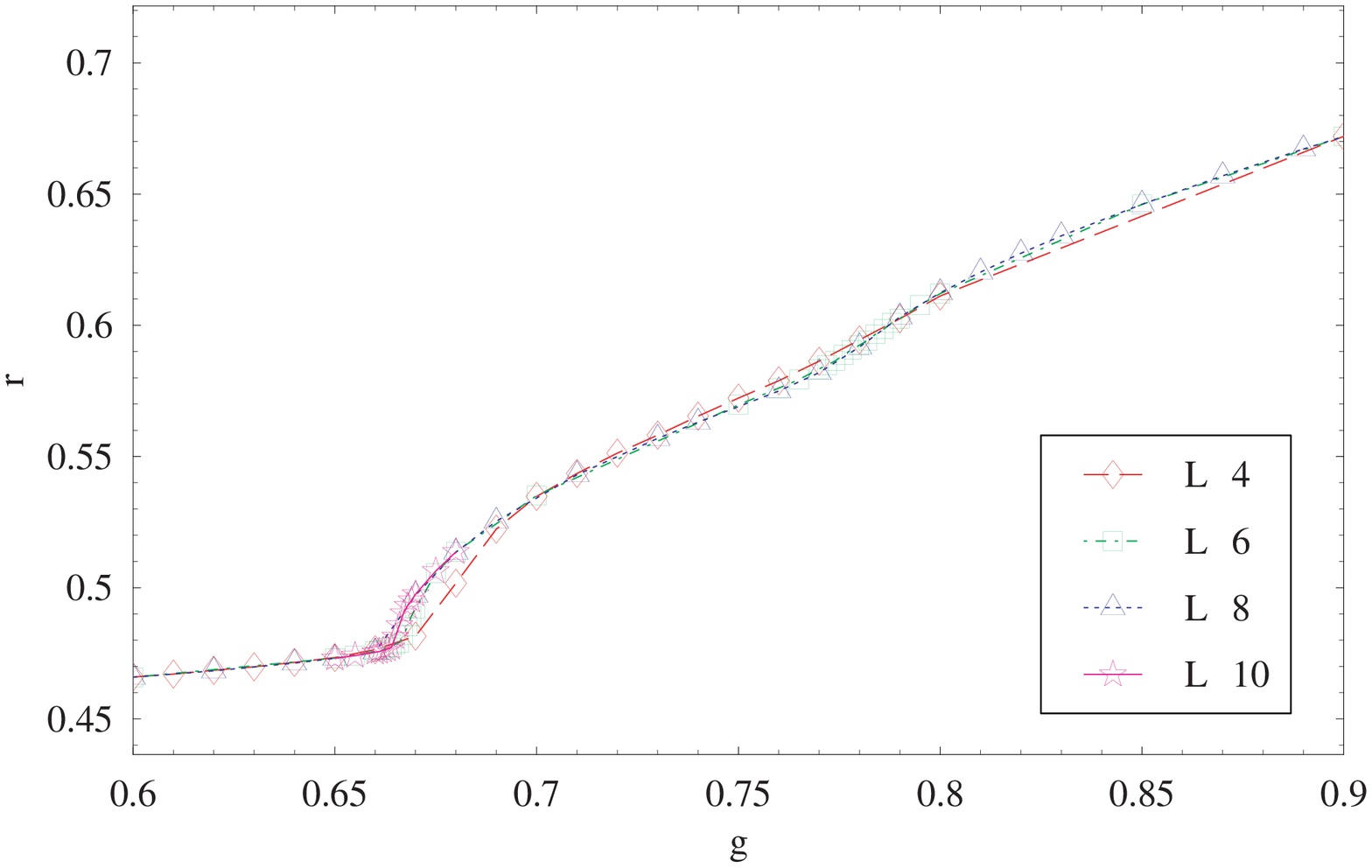}
\epsfxsize=6cm
\epsffile{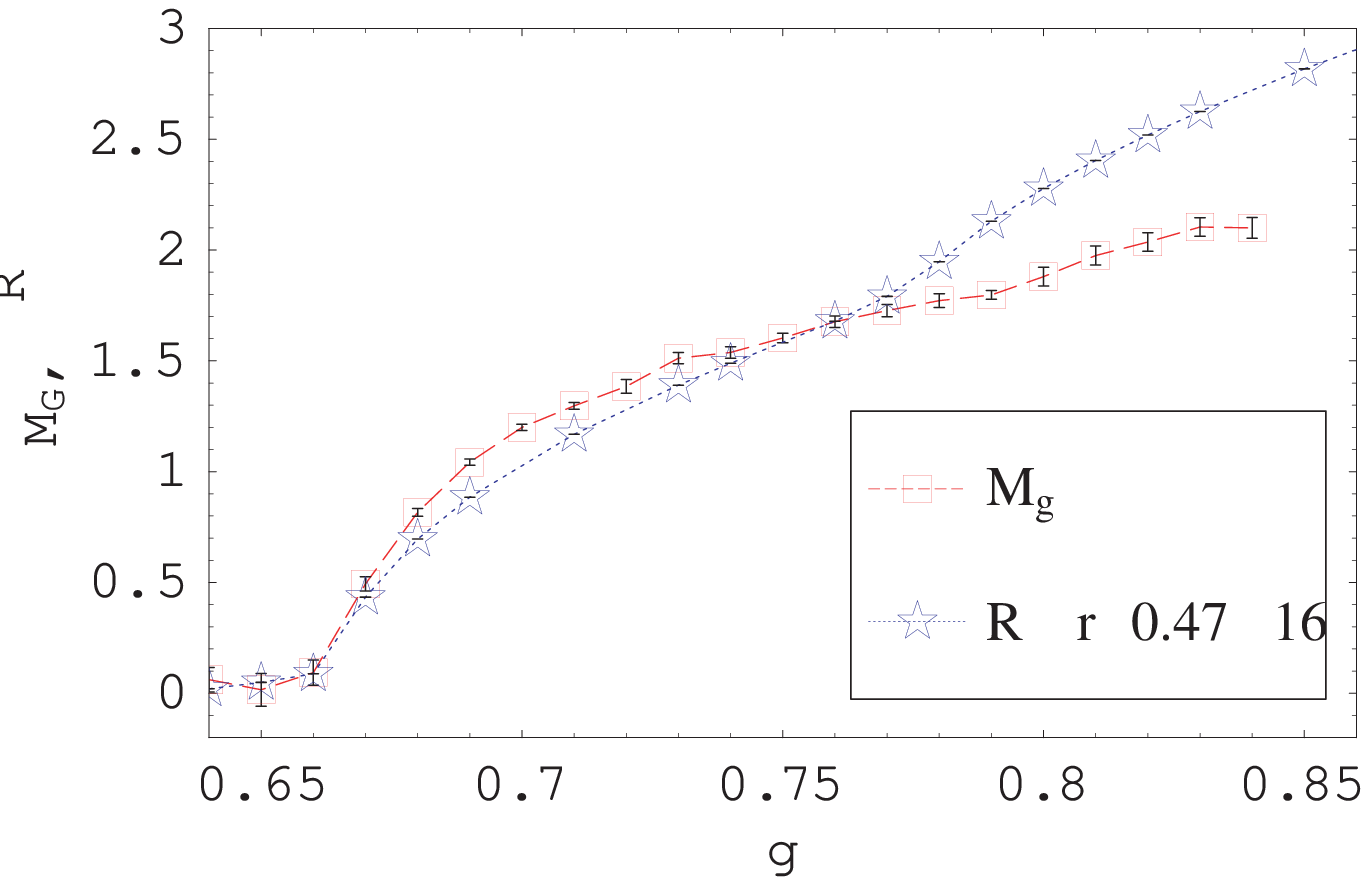}
\caption{(color online)
$r=\langle r_{x,i} \rangle$ for $c_3=1, \alpha=\beta=5$(left).
$M_{\rm G}$ and $r$ as a function of $g$(right).
}
\label{fig:55_1rm}
\end{center}
\end{figure}

The system with $c_2=-d_2=c_3=1$ has a similar phase structure
to that of the London limit with $c_2=-d_2=c_3=1$, i.e., it has 
two second-order phase transition points at $g_c=0.665$ and $g'_c=0.778$.
The detailed results have been given in Ref.\cite{UV3}. 
See Figs.\ref{fig:28} and \ref{fig:55_1rm}
for $C$, $r=\langle r_{x,i} \rangle$ and the gauge-boson mass $M_{\rm G}$.
The superfluid density $r=\langle r_{x,i} \rangle$ changes its behavior
at both critical couplings.
The mean-field approximation predicts $(r-r_0) \propto M_{\rm G}$,
where $r_0=\langle r_{x,i} \rangle$ for $g<g_c$.
The data in Fig.\ref{fig:55_1rm} shows the small but finite deviation 
from $(r-r_0) \propto M_{\rm G}$ due to the quantum fluctuations of $V_{x,i}$
and $A_{x,i}$.

\begin{figure}[htbp]
\begin{center}
\epsfxsize=6cm
\epsffile{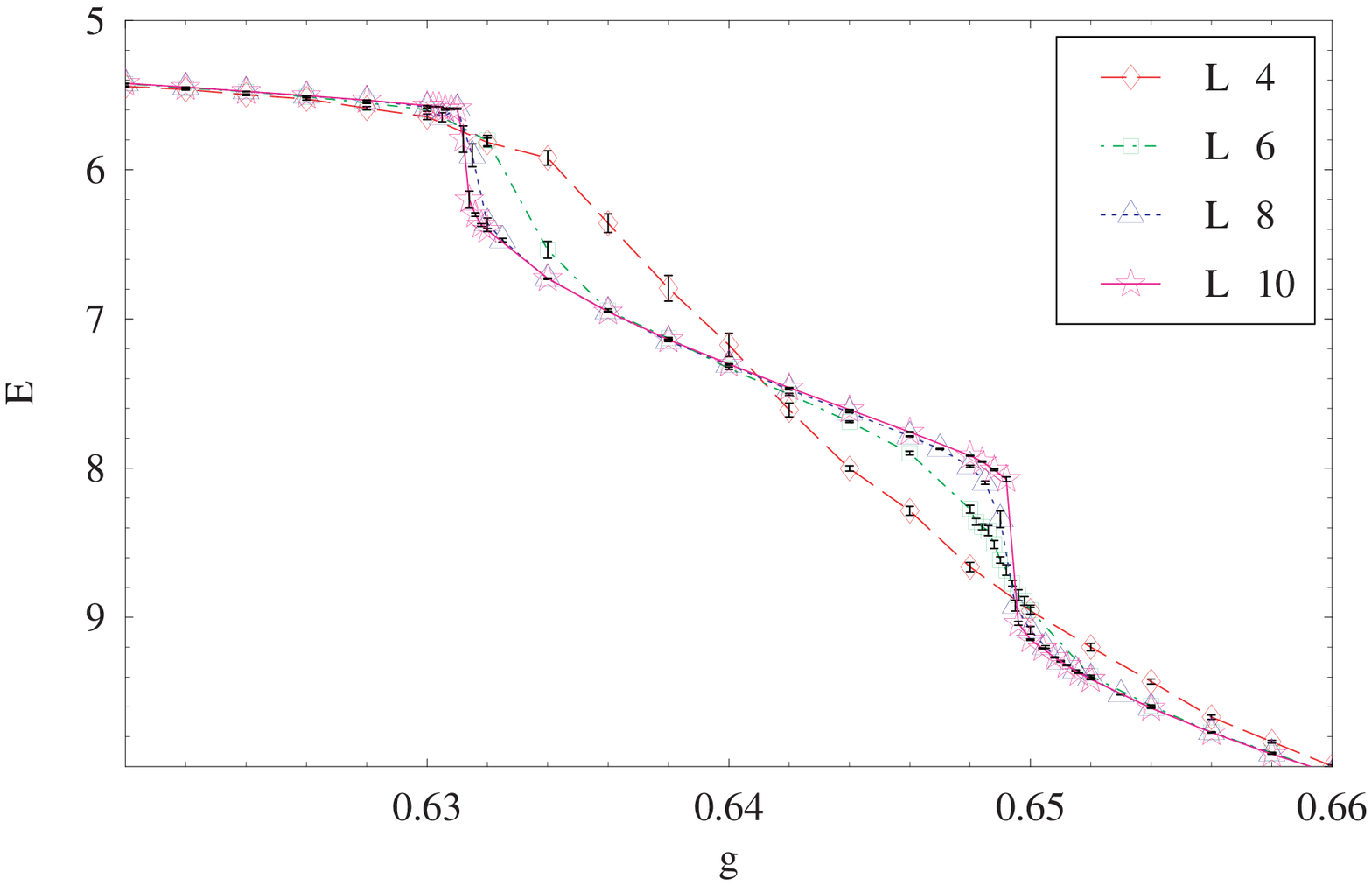}
\epsfxsize=6cm
\epsffile{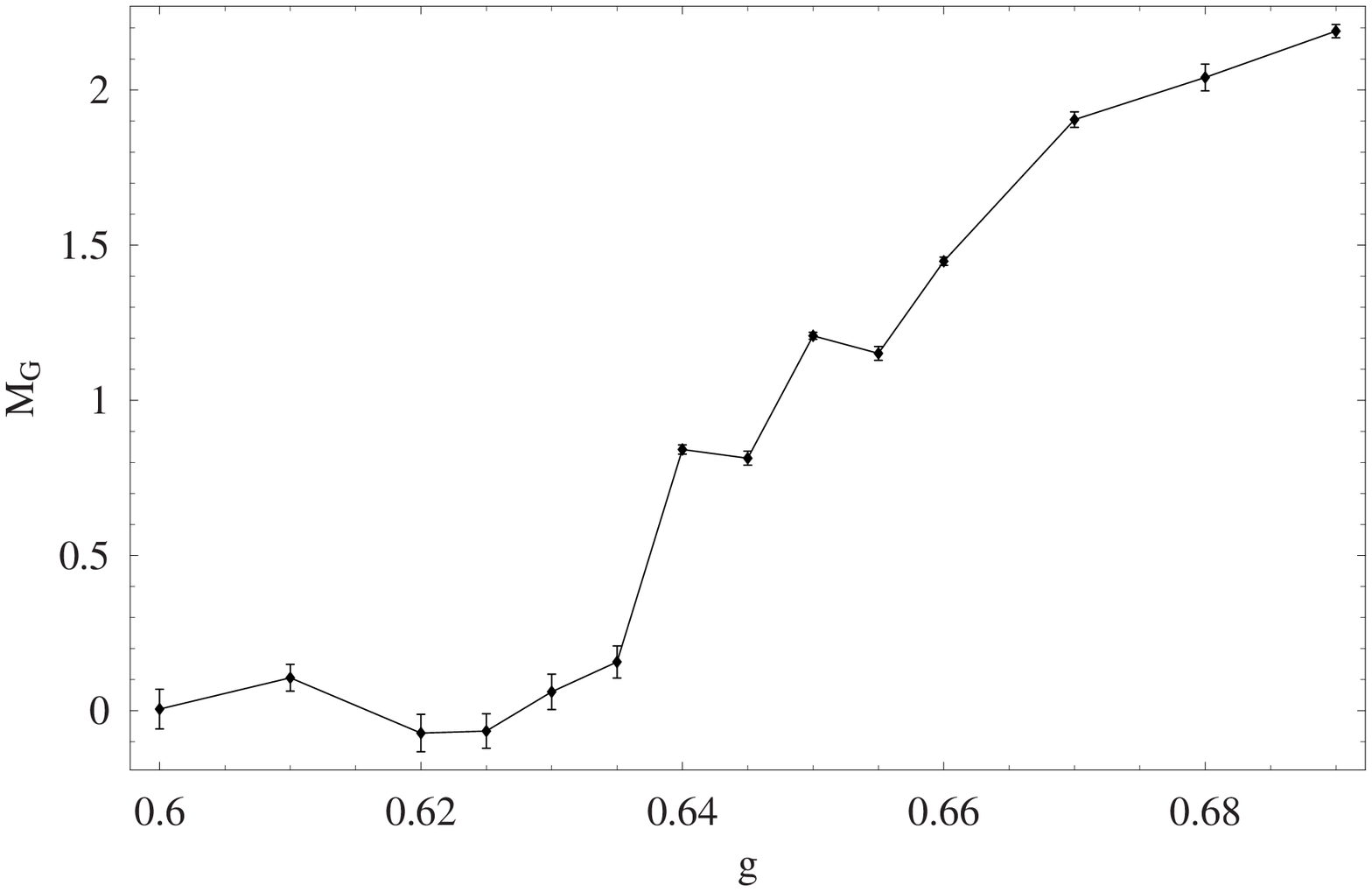}
\caption{(color online)
Internal energy $E$ and gauge-boson mass $M_{\rm G}$ for 
$\alpha=\beta=5$ and $c_3=2$.
The data of $E$ indicate the existence of {\em two first-order} 
phase transitions at $g_c=0.63$ and $g_c'=0.65$.
$M_{\rm G}$ develops at the first phase transition at $g_c=0.63$,
the SC transition, whereas it shows no clear change in behavior at
$g_c'=0.65$.
}
\label{fig:55_2E}
\end{center}
\end{figure}

\begin{figure}[htbp]
\begin{center}
\epsfxsize=6cm
\epsffile{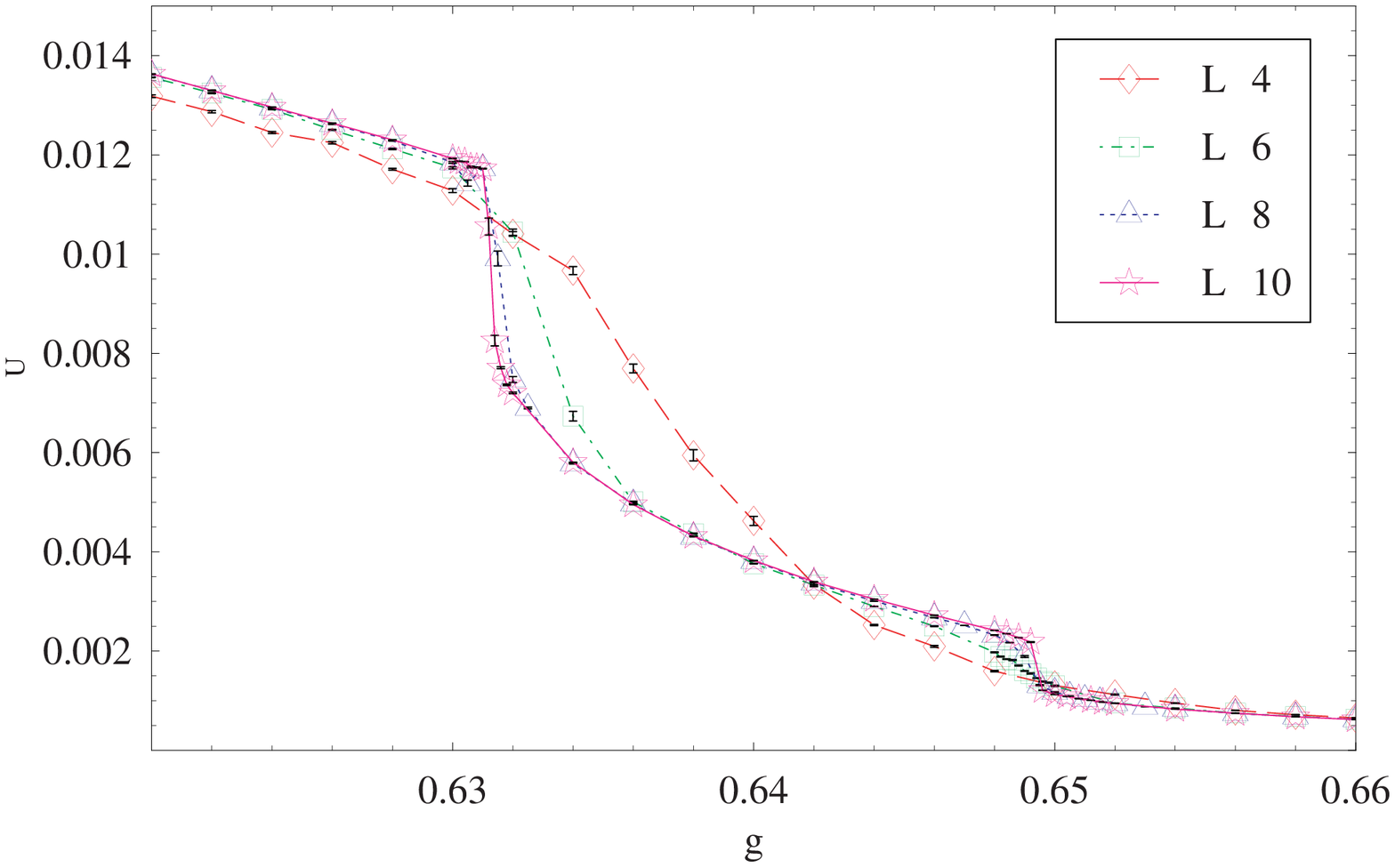}
\epsfxsize=6cm
\epsffile{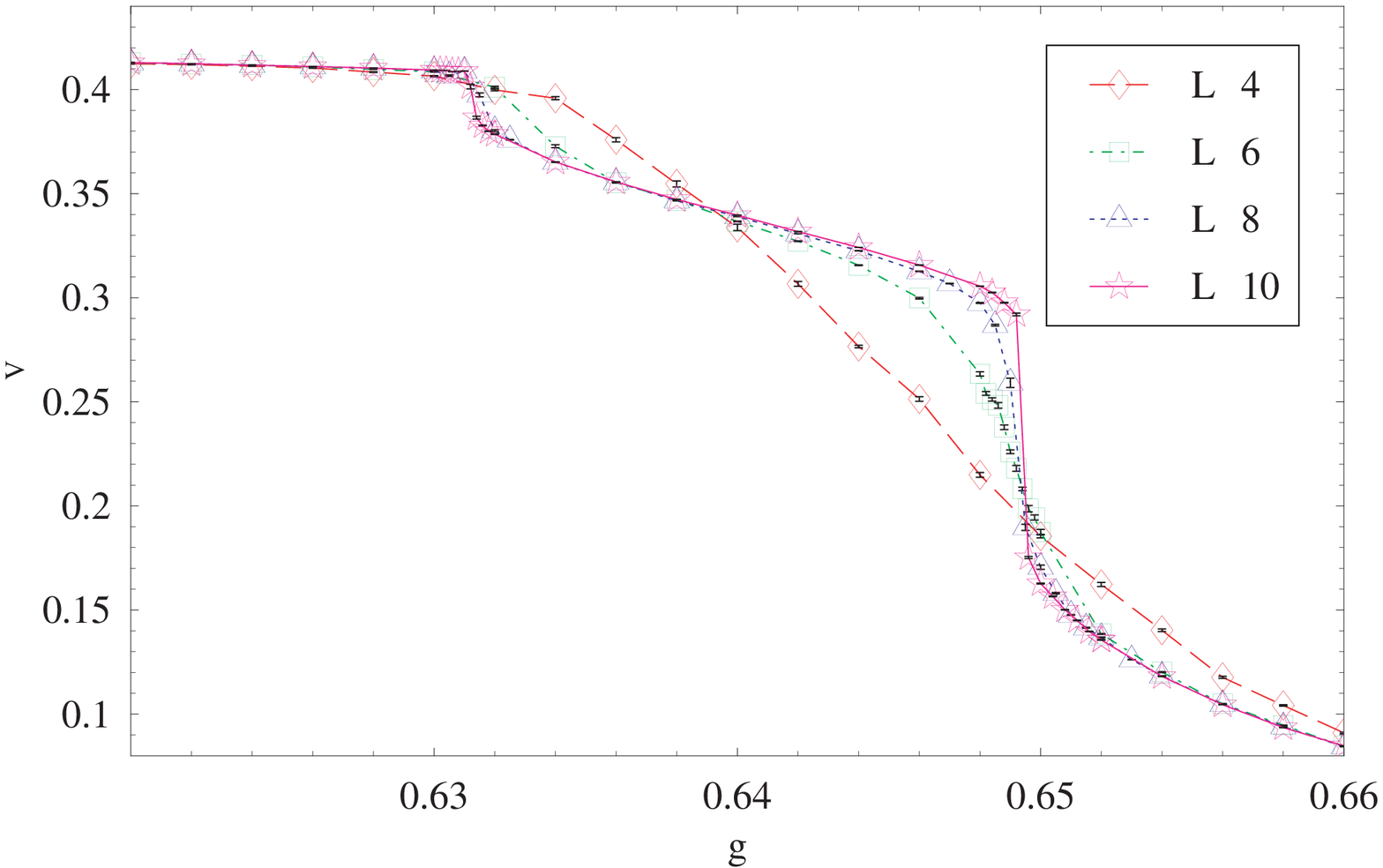}
\caption{(color online)
$\rho_U$ and $\rho_V$ for $\alpha=\beta=5$ and $c_3=2$.
The data support the existence of {\em two first-order} 
phase transitions at $g_c=0.63$ and $g_c'=0.65$.
}
\label{fig:55_2rho}
\end{center}
\end{figure}

In the rest of this section, we shall consider the cases with 
$c_3=2$ and $c_3=4$.
We first show the internal energy and $M_{\rm G}$ for the case of 
$c_2=-d_2=1$ and $c_3=2$ in Fig.\ref{fig:55_2E}.
Calculation of $E$ indicates that there exist {\em two first-order}
phase transitions at $g_c=0.63$ and $g_c'=0.65$.
We have not observed this kind of behavior for the system in the 
London limit.
$M_{\rm G}$ starts to develop at $g_c$ and therefore the first
transition at $g_c$ is the SC phase transition.
In order to verify the above conclusion that two first-order phase transitions
exist, we measured $\rho_U$ and $\rho_V$.
See Fig.\ref{fig:55_2rho}.
The both quantities exhibit sharp discontinuity at the critical couplings
$g_c$ and $g_c'$ obtained by $E$.
However, the discontinuity of $\rho_U$ ($\rho_V$) at $g_c$ ($g_c'$)
is larger than that at $g_c'$ ($g_c$).
In the region between $g_c$ and $g_c'$, $\rho_V$ has finite values as in the
previous cases.

\begin{figure}[htbp]
\begin{center}
\epsfxsize=6cm
\epsffile{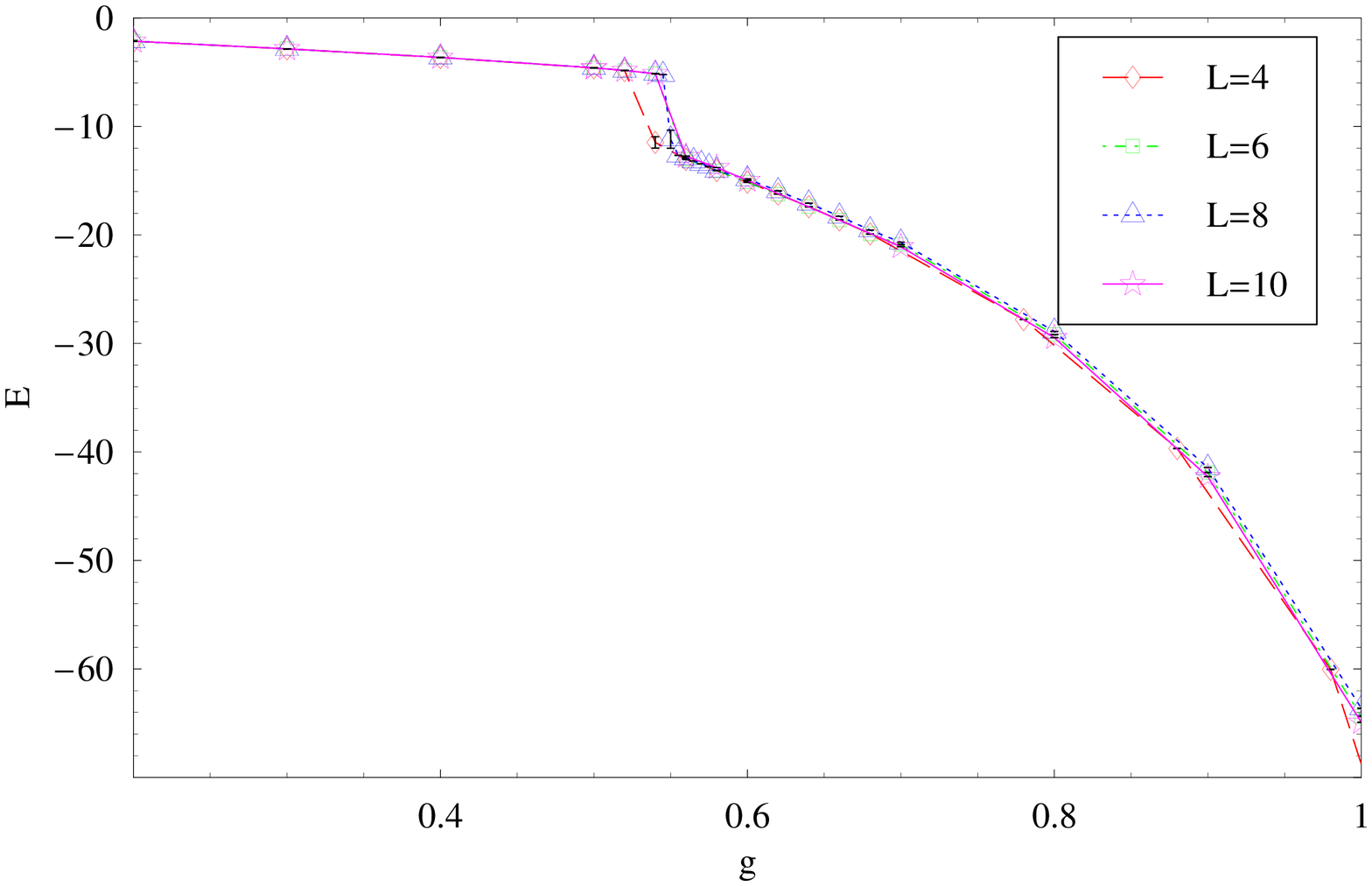}
\epsfxsize=6cm
\epsffile{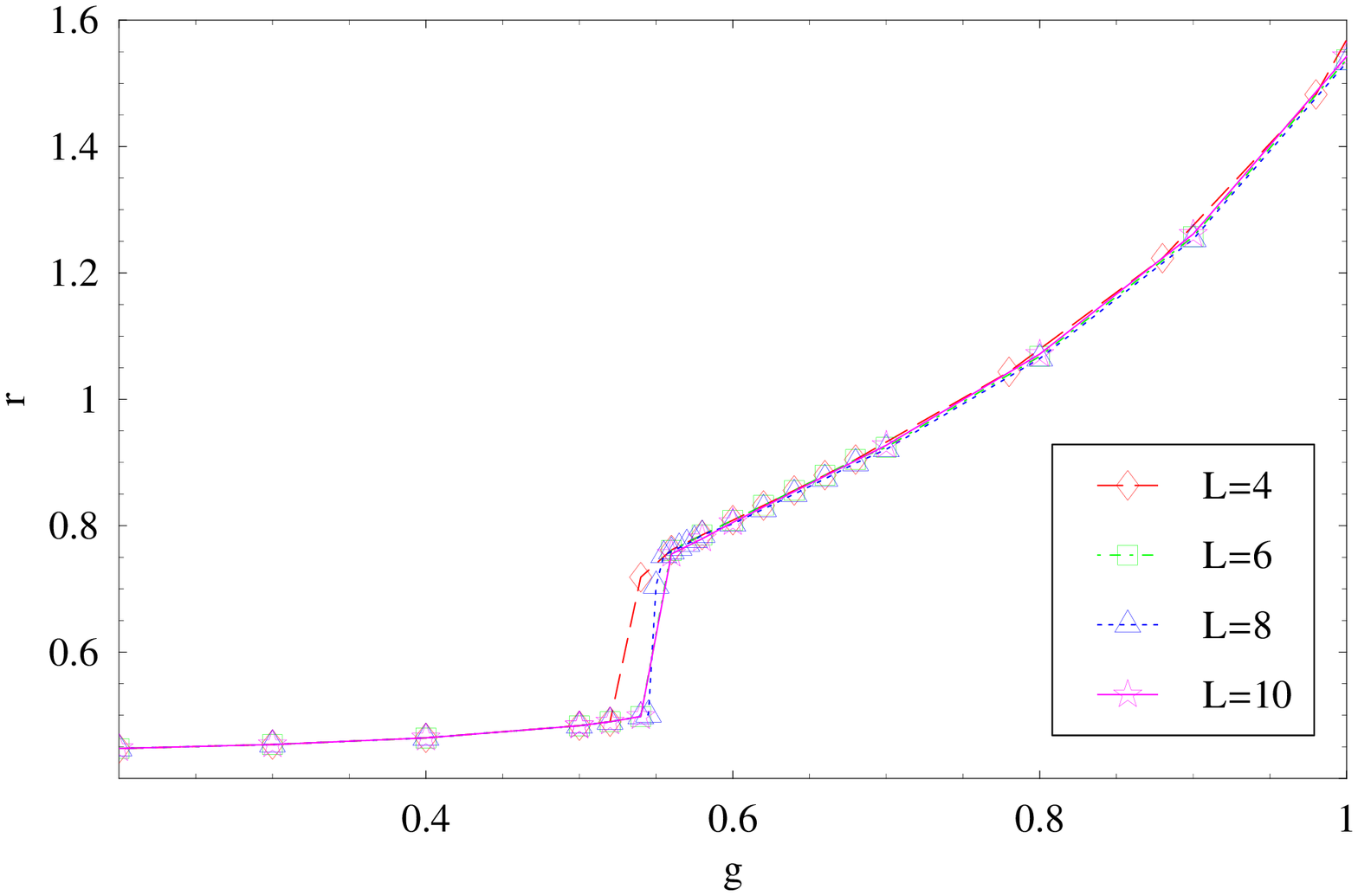}
\epsfxsize=6cm
\epsffile{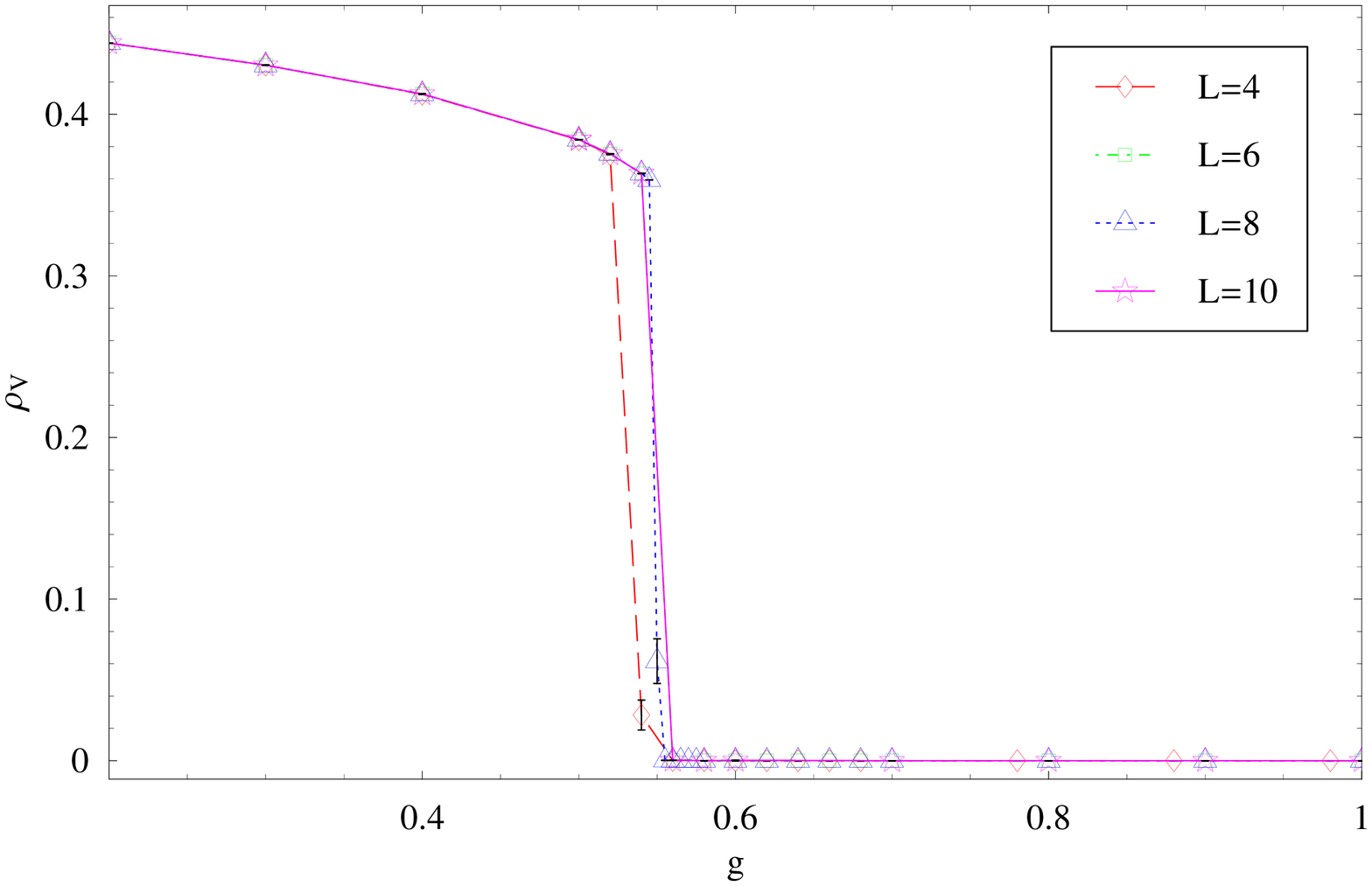}
\caption{(color online)
$E$, $r=\langle r_{x,i}\rangle$ and $\rho_V$ for $\alpha=\beta=5$ and $c_3=4$.
The data support single first-order phase transition at $g_c=0.55$.
}
\label{fig:55_4Er}
\end{center}
\end{figure}

Let us turn to the $c_3=4$ case.
$E$ in Fig.\ref{fig:55_4Er} shows that there is single first-order 
phase transition as in the London limit.
$r=\langle r_{x,i} \rangle$ and $\rho_V$ also show a discontinuity at 
$g_c=0.55$.
In the present case, two first-order phase transitions at $c_3=2$ merge into
a single first-order phase transition.

\section{Conclusion}
\setcounter{equation}{0}
In this paper we have studied the phase structure and critical behavior of 
the GL theory that we proposed earlier.
In particular, we focused on the isotropic 3D case and investigated
its quantum phase structure.
In the previous paper, we found that the 2D GL system has a second-order
SC phase transition that is in accordance with the experiments.

Though most of the materials of unconventional $d$-wave SC's have layered 
structure, properties of the phase transitions are governed by their 
three-dimensionality, e.g., the finite-$T$ phase transition is possible
by the three-dimensionality of the materials.
Then we studied the 3D system and found that interesting phase structure
appears depending on the magnitude of the parameters.
For materials with strong three-dimensionality, 
the results obtained in the present paper predict one of the following 
alternative possibilities;
\begin{enumerate}
\item
There exists CP monopole proliferation (suppression) phase transition 
within the SC phase 
\item
Single first-order SC phase transition exists accompanying monopole
transition of the CP
\end{enumerate}
As the real material of high-$T_c$ SC have strong anisotropy,
signal of the phase transition within the SC phase and first-order
phase transition might be too weak to be observed.
However the prediction itself is very interesting.

In the GL theory obtained from the canonical microscopic
model of the high-$T_c$ SC
like the t-J model, the parameter $c_3$ is an increasing function 
of the antiferromagnetic (AF) exchange coupling $J$, the amplitude of the
resonating-valence-bond configuration and the hole concentration\cite{IM}.
Then it is interesting and also important to take into account the
effects of the AF background, i.e., the coefficients in $S_{\rm GL}$ 
are not simple parameters but are determined by the dynamics of the
AF background.
This problem is under study and the result will be reported in a future
publication.

Another interesting problem is the relation between the magnetic
penetration depth at $T=0$ (the inverse gauge-boson mass at $T=0$)
and the SC phase transition temperature $T_c$.
In principle, $T_c$ can be calculated as a function of $g$ in the 
present GL theory by means of MC simulations.
Then the exponent $\alpha$, $T_c \propto (M_{\rm G})^\alpha$, should
be compared with the experimental data of the high-$T_c$ cuprates.
(Experiments give $\alpha \sim {1 \over 2}$\cite{exp}.)
This problem is also under study and the result will be published
in a future publication.


\end{document}